\documentclass[useAMS,usenatbib,fleqn]{mn2e}
\usepackage{graphicx, amssymb}
\usepackage[fleqn]{amsmath}
\usepackage{epsfig}
\usepackage{color}
\usepackage{epstopdf}
\usepackage{times}

\usepackage{enumerate}

 \title[Galactic Disk Chemistry]{Chemistry and Radiative Shielding in Star Forming Galactic Disks}
        
\author[C. Safranek-Shrader et al.]
{Chalence~Safranek-Shrader$^{1,4}$\thanks{E-mail: ctss@berkeley.edu}, Mark R. Krumholz$^{1,2}$, Chang-Goo~Kim$^3$, \newauthor 
Eve C. Ostriker$^3$, Richard I. Klein$^{4,5}$, Shule~Li$^4$, Christopher F. McKee$^{4,6}$, James M. Stone$^3$ \\
$^1$Department of Astronomy and Astrophysics, University of California, Santa Cruz, CA 95064, USA\\
$^2$Research School of Astronomy and Astrophysics, Australian National University,
Canberra, ACT 2611, Australia \\
$^3$Department of Astrophysical Sciences, Princeton University, Princeton, NJ 08544, USA \\
$^4$Astronomy Department, University of California, Berkeley, CA 94720, USA \\ 
$^5$Lawrence Livermore National Laboratory, P.O. Box 808, L-23, Livermore, CA 94550, USA\\ 
$^6$Physics Department, University of California, Berkeley, CA 94720, USA \\ 
}

\newcommand{\nh}{n_{\mathrm{\textsc{h}}}}
\newcommand{\kelvin}{\mathrm{K}}
\newcommand{\cc}{\mathrm{cm}^{-3}}
\newcommand{\msun}{M_{\odot}}

\newcommand{\htwo}{{\rm H_2}}

\newcommand{\cs}{c_{\mathrm{s}}}

\newcommand{\lj}{L_{\mathrm{J}}}

\newcommand{\av}{A_{\rm V}}
\newcommand{\co}{ \textsc{co}}
\newcommand{\despotic}{\textsc{despotic}}

\newcommand{\vect}[1]{\boldsymbol{#1}}

\begin{document}

\label{firstpage}

\maketitle
\topmargin-1cm

\begin{abstract}
To understand the conditions under which dense, molecular gas is able to form within a galaxy, we post-process a series of three-dimensional galactic-disk-scale simulations with ray-tracing based radiative transfer and chemical network integration to compute the equilibrium chemical and thermal state of the gas.  In performing these simulations we vary a number of parameters, such as the ISRF strength, vertical scale height of stellar sources, cosmic ray flux, to gauge the sensitivity of our results to these variations. Self-shielding permits significant molecular hydrogen ($\htwo$) abundances in dense filaments around the disk midplane, accounting for approximately $\sim10-15$\% of the total gas mass. Significant CO fractions only form in the densest, $\nh\gtrsim10^3\,\cc$, gas where a combination of dust, $\htwo$, and self-shielding attenuate the FUV background. We additionally compare these ray-tracing based solutions to photochemistry with complementary models where photo-shielding is accounted for with locally computed prescriptions. With some exceptions, these local models for the radiative shielding length perform reasonably well at reproducing the distribution and amount of molecular gas as compared with a detailed, global ray tracing calculation. Specifically, an approach based on the Jeans Length with a $T=40\,\kelvin$ temperature cap performs the best in regards to a number of different quantitative measures based on the $\htwo$ and CO abundances.
\end{abstract}

\begin{keywords}
stars: formation --- stars: mass function
\end{keywords}

\section{Introduction}
\label{sec:intro}

The interstellar medium (ISM) of a typical disk galaxy is divided into variety of distinct phases, generally classified into hot ionized ($T\gtrsim10^5\,\kelvin$) gas, the warm neutral medium (WNM; $T\sim10^4\,\kelvin$), and the cold neutral medium (CNM; $T\lesssim100\,\kelvin$). Star formation, however, appears to be restricted to cold dense gas with a star formation rate (SFR) that strongly correlates with the molecular gas content \citep{Wong02,Kennicutt07,Bigiel08,Leroy08,Saintonge11a,Saintonge11b,Leroy13a}. This link between star formation and molecular gas is probably not causal, in the sense that the presence of molecules simply marks those parts of the ISM that are cold and dense enough to undergo gravitational collapse \citep{Krumholz11b, Krumholz12e, Glover12a, Glover12b}, but the molecular gas
nonetheless represents an indispensable tracer of star forming regions in the local and distant Universe. Understanding these crucial links between star formation, the ISM, and galaxy evolution, all intricately coupled through a variety of energetic stellar feedback mechanisms, must rely on identifying the pathways and conditions under which cold, molecular gas is able to develop. Furthermore, the gas temperature sets the characteristic size ($R\propto T$) and mass ($M\propto T^2$) of prestellar cores that develop in post-shock regions within magnetized, star-forming clouds \citep[e.g.,][]{Chen14,Chen15}. Since gas cooled by rotational transitions of CO is able to reach lower temperature than that cooled by fine-structure lines of atomic carbon, tracking molecular chemistry is crucial to accurately represent small-scale fragmentation in numerical simulations of star-forming clouds.

Young, massive stars are the main source of far-ultraviolet (FUV) photons that permeate the ISM and readily photodissociate interstellar molecules, such as molecular hydrogen ($\htwo$) and carbon monoxide (CO). Molecular gas is thus found predominantly in dense, cold regions where dust shielding and self-shielding by the molecules themselves have attenuated the interstellar radiation field (ISRF) intensity far below its mean. The transition zones separating atomic and molecular gas, so called photodissociation regions (PDRs), have been studied extensively both numerically and theoretically \citep[e.g.,][]{Federman79,vanDishoeck86,Sternberg88,Draine96,Hollenbach99,Krumholz08,Wolfire10,Sternberg14a,Bialy15a}. Typically, the outer, unshielded layers of a PDR are composed of atomic hydrogen and singly ionized carbon. As column density increases the atomic-to-molecular transition begins to occur, with the C-CO transition occurring interior to the ${\rm H\,\textsc{i}}$-$\htwo$ one. Deeper still, when external FUV photons have been entirely absorbed, the chemistry and thermodynamics becomes dominated by the presence of deep penetrating cosmic rays and the freeze out of molecules onto dust grains.

Simply understanding the structure of a non-dynamic, one-dimensional PDR is of limited utility. The ISM is a dynamic environment where a variety of non-linear, coupled processes determine the global distribution and state of gas. Star forming complexes are continually assembled through self-gravity, while stellar feedback acts disperse the structures. Turbulent motions, ubiquitous in the ISM, dissipate through radiative shocks, but are continually replenished through a combination of supernovae and stellar winds. Thermal processes, such as dust grain photoelectric heating and molecular line cooling, further influence the state of the ISM and are sensitive functions of ambient radiation fields, column density, and the non-equilibrium chemical state of the gas. 

Implementing these processes in a numerical simulation is highly non-trivial, a challenge further exacerbated by the enormous range of spatial scales involved; the star forming region of a galactic disk can extend radially for tens of kiloparsecs while individual star forming disks are typically hundreds of AUs and smaller, making this a computational \emph{tour de force} for even adaptively refined grids. Nevertheless, three-dimensional simulations of finite, but representative, portions of galactic disks have attempted to replicate a supernova-driven ISM, finding success in reproducing a multiphase ISM \citep{Joung06,Hill12,Gent13}, or demonstrating the regulation of the SFR by stellar feedback \citep{Hopkins11a,Shetty12,Creasey13,Kim11,Kim13,Hennebelle14}.

Studies that have examined the atomic-to-molecular transition within large-scale, three-dimensional simulations has received far less attention. \citet{Glover07a,Glover07b} and \citet{Mac-Low12a} simulated the conversion of the atomic ISM to molecular form in isolated periodic boxes, without explicit feedback but including driven or decaying turbulence. \cite{Smith14} and \cite{Dobbs08} included non-equilibrium $\htwo$ formation in simulations of galactic discs, but neglected self-gravity and supernova feedback. These studies did include an $\htwo$-dissociating FUV photo-background, though assumed it to be constant in both space and time. \citet{Walch14} recently conducted a series of galactic scale simulations including supernova feedback, non-equilibrium chemistry, and radiative shielding based on the TreeCol algorithm \citep{Clark12}. Here, the strength of the FUV photodissociating background scales linearly with the instantaneous SFR, though is spatially uniform. \citet{Walch14} finds the amount and distribution of molecular gas depends sensitively on a number of parameters, in particular the precise rhythm and spatial distribution of supernovae. 

While these simulations have provided insight into the relationship between dynamics and chemistry, the price of simulating 3D time-dependent chemistry is that their treatment of photodissociation, the crucial process for regulating the chemical state of the ISM, is extremely primitive. The true photodissociation and photoheating rate at any point depends on the flux of FUV radiation integrated over all solid angles and from sources at all distances. In contrast, most of the chemodynamical simulations conducted to date have relied on \textit{ad hoc} prescriptions for the dust- and self-shielding-attenuated photodissociation rates in which a single, uniform, permeating radiation field is assumed to exist in all unshielded regions, and the degree of attenuation at a point is characterized by a single characteristic column density. These prescriptions are for the most part untested, and are of unknown accuracy.

Here our goal is to perform a detailed study on the chemical state of gas in a self-consistently simulated ISM, placing special emphasis on accurately computing the molecular content with an rigorous treatment of the generation and attenuation of photodissociating FUV radiation. We do this by post-processing the supernovae driven galactic disk simulations of \citet{Kim15a} with multifrequency ray tracing and a physically motivated spatial distribution of stellar sources. This approach has the price that it discards time-dependent dynamical effects. However, it does not rely on untested approximations for the radiative transfer problem. It therefore provides a useful complement to the more approximate dynamical simulations that have appeared in the literature thus far. In particular, our approach enables tests to evaluate the accuracy of simplified shielding schemes that have been previously used.

In this paper, our main goal is to compare a number of commonly used local approximations for the degree of radiative shielding to full solutions of the radiative transfer equation, in order to gauge the effect that different approaches to radiative shielding have on the molecular abundances and temperature.  For our tests, we use a density structure obtained from large-scale ISM simulations with turbulence self-consistently driven by star formation feedback, which naturally includes vertical stratification and strong density contrasts between warm and cold gas phases. We are also able to test how the distribution of molecular complexes is sensitive to parameters such as the stellar distribution and overall FUV intensity, and the relative importance of dust- and self-shielding.  Our comparison of different shielding prescriptions will advise us as to the validity and best use of local shielding approximations in upcoming multidimensional simulations of star formation that we shall perform.

We organize this paper as follows. In Section \ref{sec:method} we describe our methodology, including a description of the simulation we apply our post-processing to, the details of our chemical network, and the numerical approach to solving the equation of radiative transfer. In Section \ref{sec:results} we display our results. Finally, we discuss our results and conclude in Section \ref{sec:discussion}.
 
\section{Methodology}
\label{sec:method}

In this paper we apply radiative and chemical post-processing to the large-scale galactic disk simulations of \citet{Kim13} to compute the equilibrium chemical and thermal state of the disk given a realistic distribution of stellar sources, a self-consistently derived morphological structure, and multi-frequency, long-characteristics radiative transfer. 

Our methodology broadly consists of two parts. First, we perform frequency dependent radiative transfer through the simulation volume, which we describe in detail in Section \ref{sec:raytrace}, supplying us with the shielding attenuated chemical photorates. Next, these photorates are passed to a chemical network, described in Section \ref{sec:chemistry}, which is then run to equilibrium, on a cell-by-cell basis, to produce a three-dimensional datacube of chemical abundances and, in a subset of our models, gas temperature (Section \ref{sec:thermal}). In this procedure, the ray trace and chemical network integration are intricately linked via the photodissociation rates. The chemical abundances have an effect on the global photodissociation rates, which in turn determine the chemical abundances. Given the non-local coupling between these two steps, chemical network integration and ray tracing must be carried out iteratively until convergence is obtained in the chemical abundances.

\subsection{Self-consistent galactic disk simulations}
\label{sec:galactic_disk}

The simulations of \citet{Kim15a} (hereafter K15) were run with the \textsc{athena} code \citep{Stone08} which solves the equations of ideal magnetohydrodynamics on a uniform grid. The models also include self- and external-gravity, heating and cooling, thermal conductivity, Coriolis forces, and tidal gravity in the shearing box approximation to model the effect of differential galactic rotation. While K15 ran a suite of simulations with a range of magnetic field strengths and configurations, here we focus exclusively on their MB10 (solar neighborhood analog) model that has the following initial properties: gas surface density of $\Sigma_{\rm gas} = 10 \,\msun \,{\rm pc}^{-2}$, midplane density of stars plus dark matter $\rho_{\rm sd} = 0.05\,\msun \, {\rm pc}^{-3}$, galactic rotation angular speed $\Omega=28\,{\rm km} \,{\rm s}^{-1}\,{\rm kpc}^{-1}$, uniform azimuthal magnetic field of $|B_y| = 1\,\mu G$, and box size $L_x = L_y = 512\, {\rm pc}$, $L_z = 1024 \,{\rm pc}$, where the $z$ direction is perpendicular to the disk. The computational box had a uniform resolution of $dx=2\,{\rm pc}$ in each direction such that the number of grid cells along each direction were $N_x = N_y = 256$ and $N_z=512$.

Gas cooling, dominated by Ly$\,\alpha$ emission at high temperatures and ${\rm C}\,\textsc{ii}$ fine structure emission at lower temperatures, is included by use of the fitting formula of \citet{Koyama00},
\begin{align}
\Lambda(T) = \,&2 \times10^{-19} \,{\rm exp}\left(\frac{-1.184\times10^5}{T+1000}\right) \nonumber \\ 
&+2.8\times10^{-28}\sqrt{T}\,{\rm exp}\left(\frac{-92}{T}\right) \,{\rm erg}\, {\rm cm}^3\, {\rm s}^{-1}\,,
\label{eq:cooling}
\end{align}
where $T$ is the gas temperature in Kelvin. Treating radiation from young, massive stars to be the main source of heating via the dust grain photoelectric effect, the gas heating rate is set to be proportional to the recent, global SFR surface density $\Sigma_{\rm SFR}$, 
\begin{equation}
\Gamma = \Gamma_0\left[\left(\frac{\Sigma_{\rm SFR}}{\Sigma_{\rm SFR,0}}\right) + \left(\frac{J_{\rm FUV,meta}}{J_{\rm FUV,0}}\right)\right]\,{\rm erg}\, {\rm s}^{-1}\,,
\label{eq:heating}
\end{equation}
where $\Gamma_0 =2\times10^{-26} \,{\rm erg}\, {\rm s}^{-1} $ is the solar neighborhood heating rate and $\Sigma_{\rm SFR,0} = 2.5\times10^{-3}\msun \,{\rm kpc}^{-2}\,{\rm yr}^{-1}$ is the SFR surface density in the solar neighborhood. The term $J_{{\rm FUV,meta}}/J_{{\rm FUV,}0} = 0.0024$ accounts for a metagalactic FUV radiation field. Radiative shielding of FUV photons, an important effect at high gas column densities, is not included.

Supernova feedback is included in the form of instantaneous momentum injection at a rate proportional to the local star formation rate, as set by the free-fall time of gas with an efficiency of $1\%$ above a critical density threshold. The momentum from each supernova is set to $p_* = 3\times10^5\msun \,{\rm km}\, {\rm s}^{-1}$ representing the value at the radiative stage of a single supernova remnant (see \citealt{Kim15} and references therein). This approach does not form or follow the hot ($T\gg10^5\,\kelvin$) ISM phase, but instead focuses on the self-consistent turbulent driving, dissipation, and gravitational collapse of the atomic medium, the main gas reservoir of the ISM. The absence of a hot phase will have only a minimal effect on our results, since we are mainly concerned with molecule formation processes that are restricted to much lower temperatures.

\subsection{Non-equilibrium chemical network}
\label{sec:chemistry}

We utilize the chemical network described in \citet{Glover10} which we briefly summarize here. The full chemical network consists of 32 atomic and molecular species and $218$ chemical and photoreactions. The abundances of only 14 of these species (H$^+$, $\htwo$, He$^+$, C$^+$, O$^+$, OH, H$_2$O, CO, C$_2$, O$_2$, HCO$^+$, CH, CH$_2$, and CH$_3^+$) are computed by formal integration, i.e., via solving a stiff set of coupled ordinary differential equations. The abundances of the remaining species are computed by either assuming instantaneous chemical equilibrium (as in the case of H$^-$, $\htwo^+$, H$_3^+$, CH$^+$, CH$_2^+$, OH$^+$, H$_2$O$^+$, and H$_3$O$^+$, species that react so rapidly as to always be close to their equilibrium abundance values) or by utilizing conservation laws (for e$^-$, H, He, C, and O), thus reducing the number of coupled differential equations, and computational cost, that must be integrated to solve the full chemical network.

To model photodissociation and photoionization, we take the shape and strength of the interstellar radiation field (ISRF) to follow the standard \citet{Draine78} field. In Habing units, the strength of the Draine field is $G_0=1.7$, where the \citet{Habing68} radiation field is defined to be
\begin{equation}
G_0 = \frac{4\pi\int_{\rm FUV} J_{\nu}d\nu}{1.6\times10^{-3}\,{\rm erg}\,{\rm cm}^{-2}\,{\rm s}^{-1}}\,,
\label{eq:habing}
\end{equation}
where $J_\nu$ is the angle averaged specific intensity and the integral runs over the FUV range, from $6-13.6\,{\rm eV}$. The photodissociation and photoionization rate for any given chemical species can be written as
\begin{equation}
k_{\rm thick} = k_{\rm thin}\,G_0\, f_{\rm shield} \equiv k_{\rm thin}\,G_{0,{\rm eff}}\,,
\label{eq:photo_shield}
\end{equation}
where $k_{\rm thin}$ is the photorate in optically thin gas (see table B2 in \citealt{Glover10}), and $f_{\rm shield}$ is the degree by which the photorate is reduced as compared with the optically thin rate. We defer a discussion of how the shielding factors $f_{\rm shield}$ and unattenuated radiative intensity $G_0$ are determined to Section \ref{sec:raytrace}.

\subsection{Thermal processes}
\label{sec:thermal}

In a subset of our models we evolve the temperature to equilibrium along with the chemical abundances. In this section we describe these heating and cooling processes. 

In the low temperature, moderate density, molecular ISM, line emission originating from the rotational transitions of CO is a major source of radiative cooling. The volumetric cooling rate from CO can be written as $\Lambda_{\rm \textsc{co}} = L_{\rm \textsc{co}}\,n_{\rm \textsc{co}}\,\nh$. In general, the CO cooling function $L_{\rm \textsc{co}}$ is a complicated function of temperature, density, column density, and velocity dispersion, and in a multidimensional simulation it is not feasible to self-consistently solve for the level populations in every computational cell. 

Hence, we employ the \despotic~code \citep{Krumholz14} to pre-compute the CO cooling function. Utilizing the large velocity gradient approximation, we tabulate $L_{\rm \textsc{co}}$ as a function of $n_{\htwo}$, temperature, and, following \citet{Neufeld95}, an effective column density per velocity $\tilde{N}_{\rm \textsc{co}}$ defined such that
\begin{align}
\tilde{N}_{\rm \textsc{co}} = \frac{n_{\rm \textsc{co}} }{ |\nabla \cdot \vect{v}|}\,,
\label{eq:ntilde}
\end{align}
where $n_{\rm \textsc{co}}$ is the CO number density. The local velocity divergence, $\nabla \cdot \vect{v}$,  is computed using a three-point stencil around the cell of interest. During the computation we read in the \despotic-computed cooling tables and employ tri-linear interpolation to compute the CO cooling rate as needed.

In addition to CO line cooling, we also consider fine-structure line emission from $[{\rm C}\,\textsc{ii}]$, $[{\rm C}\,\textsc{i}]$, $[{\rm O}\,\textsc{i}]$. In the case of these species, because their lines remain optically thin under all the conditions found in our calculations, it is straightforward to solve for the level populations and compute the cooling rate directly. We  use the collisional rate coefficients, atomic data, and methodology presented in \citet{Glover07}.

The rate of energy exchange between gas and dust is
\begin{align}
\Lambda_{\rm gd} = \alpha_{\rm gd} \, T^{1/2}\,(T-T_{\rm dust})\,\nh^2 \,,
\label{eq:dustgas}
\end{align}
where $\alpha_{\rm gd} = 3.2\times10^{-34}\,{\rm erg}\,{\rm cm}^3\,\kelvin^{-3/2}$ is the dust-gas coupling coefficient for Mliky Way dust in $\htwo$ dominated regions \citep{Goldsmith01} and $T_{\rm dust}$ is the dust temperature. In principle $T_{\rm dust}$ can be determined self-consistently by considering dust to be in thermal equilibrium between absorption and emission, but for simplicity here we assume a constant dust temperature of $T_{\rm dust}=10\,\kelvin$. This should be an acceptable approximation since gas-dust coupling does not typically become important until $n\gtrsim10^4\,\cc$, a regime not probed here.

Cooling due to the collisional excitation of atomic hydrogen, or Lyman-$\alpha$, is given by \citep{Cen92},
\begin{align}
\Lambda_{\rm Ly\alpha} = 7.5\times10^{-19}\,(1+T_5^{1/2})^{-1}\,e^{-118348/T}\,n_{\rm e}\,n_{\rm H\,\textsc{i}} \,,
\label{eq:lya}
\end{align}
where $T_5=T/(10^5\,\kelvin)$, and $n_{\rm e}$ and $n_{\rm \textsc{H}\,\textsc{I}}$ are the electron and neutral hydrogen densities, respectively.

The photoelectric heating rate, including recombination cooling, is given by \citep{Bakes94},
\begin{equation}
\begin{aligned}
\Gamma_{\rm pe} =\,&10^{-24}\epsilon \,G_{0,{\rm eff}} \nh -4.65\times10^{-30}T^{0.94}\psi^{\beta}n_{\rm e}\,\nh \,,
\label{eq:photoelectric}
\end{aligned}
\end{equation}
where $G_{0,{\rm eff}} = G_0\,{\rm exp}(-2.5\,\av)$ is the attenuated radiation field strength described in Section \ref{sec:raytrace}, $\psi = G_{0,{\rm eff}}T^{1/2}/n_e$, $\beta=0.735/T^{0.068}$, and 
\begin{align}
\epsilon = \,&\frac{0.049}{1+4\times10^{-3}\psi^{0.73}} +\frac{0.037\,(T/10^{4})^{0.7}}{1+2\times10^{-4}\psi}
\label{eq:photoelectric_eff}
\end{align}
is the photoelectric heating efficiency.

The heating rate due to cosmic ray ionization is given by
\begin{align}
\Lambda_{\rm cr} &= \zeta  \nh q_{\rm ion} \nonumber  \\
&=\zeta \nh (x_{\rm  \textsc{H}\,\textsc{i}}\,q_{ \rm ion,\textsc{H}\,\textsc{I}} + 2\,x_{\htwo}\,q_{\rm ion,\htwo})\,,
\label{eq:cosmic_ray}
\end{align}
where $\zeta$ is the cosmic ray ionization rate per hydrogen nucleus which we take to be $\zeta = 10^{-17}\,{\rm s}^{-1}$, a relatively low value that aids in the production of CO. The energy added per cosmic ray ionization $q_{\rm ion}$ depends on the chemical composition of the gas. For purely atomic gas we use the recommendation from \citet{Draine11},
\begin{align}
q_{ \rm ion,\textsc{H}\,\textsc{I}}  = 6.5 \,{\rm eV} + 26.4 \,{\rm eV}\, \left(\frac{x_e}{x_e + 0.07}\right)^{1/2}\,,
\label{eq:qionhi}
\end{align}
while we use a piecewise fit given in \citet[][equation B3]{Krumholz14} of numerical data from \citet{Glassgold12} for $q_{\rm ion,\htwo}$.

Note that we do not include any sort of heating due to turbulent dissipation or ambipolar diffusion, two related effects whose combined heating rate can be comparable to the cosmic ray heating rate for the low cosmic ray ionization rate we adopt here \citep[e.g.,][]{Li12}.

\subsection{Multi-angle ray trace and radiative shielding}
\label{sec:raytrace}

Modeling radiation transport in full generality within a three-dimensional simulation presents a significant numerical challenge. As noted by \citet{Glover10}, the cost of multifrequency radiation transfer scales as $O(N_{\nu}\times N^{5/3})$, where $N_{\nu}$
 is the number of discrete frequency bins and $N$ is the number of computational elements. Even with moderate resolution, this is orders-of-magnitude larger than the hydrodynamic evolution, which scales as $O(N)$, and generally intractable. While a number of approximate methods exist for solving the equation of radiative transfer in a numerical simulation 

 \citep[e.g.,][]{Rijkhorst06,Krumholz07b,Wise11,Davis12}, these approaches are not suitable for line transfer calculations, which are crucial for determination of the photoreaction rates. Here, we elect to solve for the radiation field, and chemical state, by direct ray-tracing applied to static simulation snapshots.

 \begin{figure}
 \begin{center}
\includegraphics[width=0.45\textwidth]{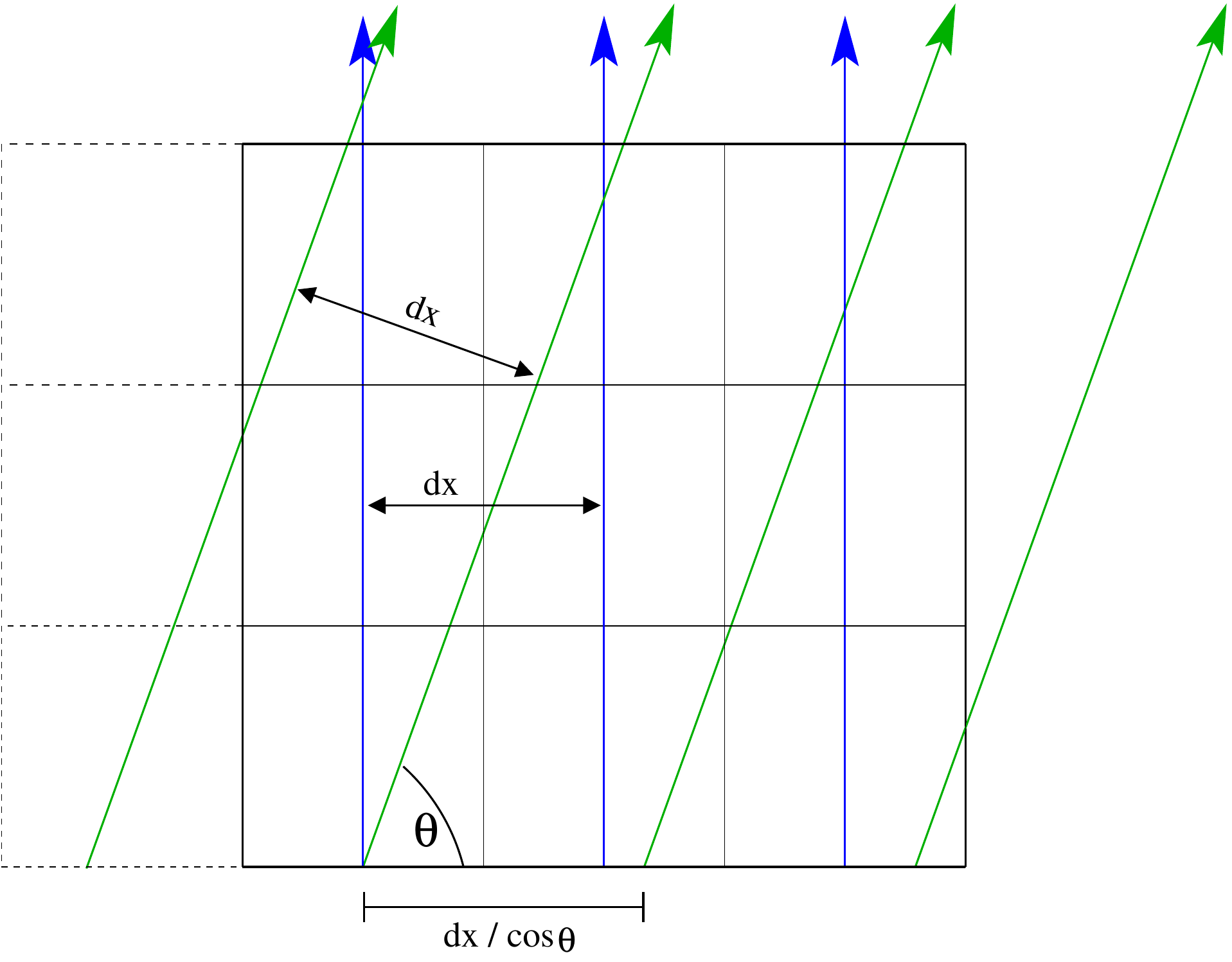}
\end{center}
\caption{Two-dimensional schematic of the ray tracing procedure. Here, the first set of rays (shown as blue arrows) are launched perpendicular to the $x$-axis with a separation equal to the cell spacing $dx$. The next set of rays (green arrows) are launched with an angle of $\theta$ with respect to the $x$-axis and maintain a perpendicular separation $dx$ with the first set of rays. This choice guarantees the computational domain is evenly sampled by rays for each angular set of rays. Only when the rays are physically located within the computational grid (solid boxes) is the ray tracing performed.}
\label{fig:gridfig}
\end{figure}

\subsubsection{Shielding factors}
\label{sec:shield}

The radiation intensity at any point along a ray is governed by the time-independent radiative transfer equation,
\begin{equation}
\frac{dI_\nu}{ds} = - \alpha_\nu I_\nu +  j_{*,\nu} \,,
\label{eq:rad_transfer}
\end{equation}
where $I_\nu$ is the specific intensity, $\alpha_\nu$ is the absorption coefficient, $j_{*,\nu}$ is the emissivity, and $s$ is the distance along the ray. We are particularly concerned with FUV photons whose dominant sources are stars with ages of $\sim 100$ Myr or less. Since the simulation does not follow young stars explicitly, and since by this age stars are generally no longer predominantly found in clusters \citep[e.g.,][]{Fall12a, Fouesneau14a}, we approximate their distribution as a smooth field that varies with height $z$ alone. We therefore assign every grid cell a frequency-dependent FUV emissivity $j_{*,\nu} = j_{*,0,\nu}f(z)$, where the normalization $j_{*,0,\nu}$ is fixed by the requirement that the unattenuated midplane ISRF intensity, $G_{0,{\rm midplane}}$, is in accordance with the current SFR surface density:
\begin{equation}
G_{0,{\rm midplane}} = 1.7\left(\frac{\Sigma_{\rm SFR}}{\Sigma_{\rm SFR,0}}\right)
\label{eq:}
\end{equation}
where $\Sigma_{\rm SFR}$ and $\Sigma_{\rm SFR,0}$ retain their definitions from Equation \ref{eq:heating}. For the vertical distribution of the FUV emissivity we adopt the profile
\begin{equation}
f(z)= \mathrm{sech}^2(z/H)\,,
\label{eq:fz}
\end{equation}
appropriate for an isothermal disc, where $H$ is the stellar scale height. For our fiducial model we adopt a scale height of $H=100\,{\rm pc}$, though we do explore how variations in $H$ affect our results.

Even in a static snapshot, solving the transfer equation in three dimensions with enough frequency resolution to resolve the dominant photodissociation lines of every important species is prohibitively expensive. Fortunately, it is also unnecessary for the purpose of computing photoreaction rates, which is our goal. Photodissociation for important interstellar species occurs via resonant absorption of an FUV photon in one or more bands that occupy a relatively narrow range of frequencies. One-dimensional simulations that include full frequency-dependent transfer show that the photodissociation rate produced by the ISRF decreases as the radiation propagates through increasing columns of material due to two main effects: continuum absorption by dust, and resonant absorption of photons near the centers of the dissociating absorption lines \citep{vanDishoeck86, Black87a, van-Dishoeck88a, Draine96, Browning03a}.

For dust, the reduction in the photodissociation rate is given by
\begin{equation}
f_{\rm shield} = f_{\rm dust} = {\rm exp}(-\gamma \av)
\label{eq:dust_shield}
\end{equation}
where $\gamma = 0.92\,\kappa_\nu / \kappa_{\rm V}$ is the scaling between the dust opacity at the photodissociation absorption frequency and the V-band. Here $\kappa_\nu$ is the dust opacity evaluated at the absorption band relevant for that species, which is taken to be at a single characteristic frequency, a good approximation since $\kappa_\nu$ varies slowly with $\nu$ over the band; $\kappa_{\rm V}$ is the V-band dust opacity. For the chemical species we consider, $\gamma$ ranges from $0.5$ to $3.9$. The visual extinction in magnitudes $\av $ is directly proportional to the total column density of H nuclei, $N_{\rm H} = N_{\rm H\,\textsc{i}} + 2N_{\htwo}$, 
\begin{equation}
\av = \frac{N_{\rm H}}{1.87\times10^{21}\,{\rm cm}^{-2}} = N_{\rm H}\,\sigma_{\rm d,V}\,,
\label{eq:av}
\end{equation}
where $\sigma_{\rm d,V} = 5.3\times10^{-22}\,{\rm cm}^{2}$ \citep{Draine96} is the total dust cross section in the V-band per hydrogen nucleus.

To properly model the photodissociation of $\htwo$ and CO we must also consider self-shielding in which the molecules themselves, in addition to dust grains, contribute to the attenuation of FUV photons. Self-shielding of $\htwo$ can be well approximated with the following analytic expression from \citet{Draine96},
\begin{align}
f_{\rm self,H_2} &= \frac{0.965}{(1+x/b_5)^2} +  \frac{0.035}{\sqrt{1+x}} \nonumber \\
  &\times {\rm exp}(-8.5\times10^{-4}\sqrt{1+x})\,,
\label{eq:fh2}
\end{align}
where $x = N_{\htwo}/(5\times10^{14}\,{\rm cm}^{-2})$, $b_5 = b/(10^5\,{\rm cm}\, {\rm s}^{-1})$, and $b=9.1\,{\rm km}\,{\rm s}^{-1}(T/10^4\,\kelvin)^{1/2}$ is the Doppler broadening parameter for $\htwo$. The total shielding factor for $\htwo$ will be $f_{\rm shield} = f_{\rm self,\htwo}\,f_{\rm dust,\htwo}=f_{\rm self,\htwo}{\rm exp}({-3.7\av})$. 

CO not only experiences self-shielding and dust shielding, but is additionally shielded by molecular hydrogen. The photodissociation rate for CO can be written as $k_{\rm thick} = k_{\rm thin} f_{\rm dust}f_{ \rm self,\co}f_{ \rm \co,\htwo}$, where $f_{ \rm self,\co} = f_{\rm self,\co}(N_{\rm \co})$ is the CO self-shielding function and $f_{\rm \co,\htwo} = f_{\rm \co,\htwo}(N_{\htwo})$ accounts for the cross-shielding of CO by the overlapping Lyman-Werner lines of $\htwo$. Both $f_{ \rm self,\co}$ and $f_{\rm \co,\htwo}$ are taken from \citet{Lee96}, while $f_{\rm dust}={\rm exp}(-2.5\,\av)$.

A critical point to realize is that, because the CO and H$_2$ self-shielding factors result from a rearrangement of the photon frequency distribution as radiation propagates along a ray, they may be applied independently on every ray. This means that, rather than requiring many frequency bins to resolve the dissociating lines of H$_2$ and CO, we can approximate the effect by considering a single frequency bin for H$_2$-dissociating photons, and similarly for CO-dissociating ones, and use the pre-tabulated shielding functions to model the reduction in photodissociation rate along a ray. The main limitation in this approach is that the shielding functions have been tabulated for gas with a fixed velocity dispersion and zero bulk velocity, whereas in our simulations the velocity fields are significantly more complex. However, we show below that our results are quite insensitive to exactly how we estimate the velocity-dispersion that enters the shielding factors.

\subsubsection{Solving the radiative transfer equation}
\label{sec:solving_rad_transfer_equation}

Based on the discussion in the preceding section, we now discretize the equation of radiative transfer (Equation \ref{eq:rad_transfer}) by defining three distinct frequency bins: one describing radiation in the FUV dust continuum, $I_{\rm cont}$, one that corresponds to the Lyman-Werner bands for $\htwo$, $I_{\htwo}$, and one describing the photodissociation of CO, $I_{\rm \textsc{co}}$. 
The three frequency bins can be formally defined through the following filters: 
\begin{align}
\theta_{\rm FUV}(\nu) &= \theta_{\htwo}(\nu) + \theta_{\rm \textsc{co}}(\nu) + \theta_{\rm cont}(\nu)\,,
\label{eq:filters}
\end{align}
defined such that $\theta_{\htwo}(\nu)$ and $\theta_{\rm \textsc{co}}(\nu)$ equal unity in the frequency ranges that overlap with their photodissociation absorption lines, and are zero otherwise. Likewise, $ \theta_{\rm cont}(\nu)=1$ in the FUV frequency range, excluding the photoabsorption lines of $\htwo$ and CO, and is zero otherwise.

The intensity at any point $s$ along a ray is given by the formal solution to Equation (\ref{eq:rad_transfer}),
\begin{equation}
I_{\nu}(s) = \int_0^s{\rm exp}(-\int_{s'}^s \alpha_{\nu}\, ds'')\, j_{*,\nu}(s')\,ds'\,.
\label{eq:soln}
\end{equation}
Identifying ${\rm exp}(-\int_{s'}^s\alpha_\nu \, ds'')$ to be the shielding factor $f_{\rm shield}$ (Equation \ref{eq:photo_shield}) and integrating over the frequency filters defined in Equation (\ref{eq:filters}) gives us the intensity in our three frequency bins:
\begin{align}
I_{\htwo}(s) &= \int_0^\infty I_{\nu}(s)\,\theta_{\htwo}(\nu)\,d\nu \nonumber \\
&=\int_0^s f_{\rm dust}(N_{\rm H}')\,f_{\rm self,\htwo}(N_{\htwo}')\,j_{*,\htwo}(s')\,ds' 
\label{eq:Ih2}
\end{align}
\begin{align}
I_{\rm \textsc{co}}(s) &= \int_0^\infty I_{\nu}(s)\,\theta_{\rm \textsc{co}}(\nu)\,d\nu \nonumber \\
&=\int_0^s f_{\rm dust}(N_{\rm H}')\,f_{\rm self,\textsc{co}}(N_{\rm \textsc{co}}')f_{\rm \textsc{co},\htwo}(N_{\htwo}')\,j_{*,{\rm \textsc{co}}}(s')\,ds' 
\label{eq:Ico}
\end{align}
\begin{align}
I_{\rm cont}(s) &= \int_0^\infty I_{\nu}(s)\,\theta_{\rm cont}(\nu)\,d\nu \nonumber \\
&=\int_0^s f_{\rm dust}(N_{\rm H}')  \,j_{\rm *,cont}(s')\,ds'\,.
\label{eq:Icont}
\end{align}
In the above equations, the shielding factors are evaluated using the total column densities between points $s$ and $s'$, $N_{\rm \{H/\htwo/\co\}}' \equiv \int_{s'}^s n_{\rm \{ \textsc{H}/\htwo/\co\}} \,ds''$, $j_{*,{\rm H_2}} = \int j_{*,\nu} \theta_{\rm H_2} \, d\nu$, and similarly for $j_{*,{\rm \textsc{co}}}$ and $j_{*,{\rm cont}}$

\subsubsection{Spatial and angle discretization}
\label{sec:spatial_discretize}

We are now in a position to discretize the Equations (\ref{eq:Ih2}), (\ref{eq:Ico}), and (\ref{eq:Icont}) in a way suitable for numerical calculation. Consider a ray passing through our Cartesian grid at some angle, and consider the $n$'th cell along that ray. Let the sequence of cells through which the ray passes on its way to that cell $n$ be numbered $i=0,1,...,n-1$, and let $\Delta s_i$ be the path length of the ray through cell $i$; let $n_{X,i}$ be the corresponding number density of species $X$ (e.g., H$_2$), and similarly for all other quantities. In this case Equation (\ref{eq:Ih2}) can be discretized to
\begin{equation}
I_{\htwo}(n) = \sum\limits_{i=0}^{n-1} \Delta s_{i}\,j_{*,\htwo,i}\,f_{\rm shield}\left(\sum\limits_{j=i}^{n}n_{\{{\rm \textsc{H}}/\htwo \},j}\Delta s_{j}\right),
\label{eq:discretize}
\end{equation}
where the shielding factor, which includes both dust- and self-shielding, is evaluated using the total column $N_{\{{\rm \textsc{H}}/\htwo \}} = \sum_{j=i}^{n}n_{\{{\rm \textsc{H}}/\htwo \},j}\Delta s_{j}$ between the cells $n$ and $i$. Similar results hold for Equations (\ref{eq:Ico}) and (\ref{eq:Icont}).

We have now written down a discretized version of the radiative transfer equation for a particular direction of propagation. The remaining step is to discretize the problem in angle. We do so by drawing rays in $N_{\rm pix}$ directions (parametrized by $\theta$ and $\phi$) selected via the \textsc{healpix} algorithm for equal area spherical discretization \citep{Gorski05}. All our models use \textsc{healpix} level $2$, corresponding $N_{\rm pix}=48$, which we have found provides sufficient resolution to properly sample the radiation field. For each direction, we draw through the computational domain a series of rays separated by one cell spacing $dx$ in the direction perpendicular to the rays, such that each ray represents a solid angle $d\Omega = 4\pi/N_{\mathrm{pix}}$ and area $dx^2$. This choice guarantees that each cell is intersected by at least $N_{\rm pix}$ rays in total. We show a schematic diagram of our ray tracing procedure in Figure \ref{fig:gridfig}. The radiation intensity in any given cell is then able to be computed through an angle average of all intersecting rays.


We are ultimately interested in photodissociation rates which are in turn proportional to the angle-averaged intensity,
\begin{equation}
k_{\rm thick,\htwo}\propto J_{\htwo} = \frac{1}{4\pi} \int I_{\htwo} \, d\Omega,
\end{equation}
and similarly for all other species. Evaluating this integral requires some subtlety; although all our rays represent an equal amount of solid angle, not all rays intersecting a cell have the same path length $\Delta s$ through it, even rays that represent the same direction of radiation propagation. The contribution to the mean intensity in a cell from a given ray $n$ will be proportional to the time that the photons propagating along that ray spend in the cell, and thus proportional to $\Delta s_n$. We therefore compute the mean intensity by weighting each ray's contribution by $\Delta s$:
\begin{equation}
J_{\htwo} \propto \left(\sum\limits_{n=1}^{N_{\rm ray}} I_{\htwo,n}\, \Delta s_n\right) \bigg/ \left(\sum\limits_{n=1}^{N_{\rm ray}} \Delta s_n\right)\,
\label{eq:soln}
\end{equation}
where $N_{\rm ray}$ is the total number of rays that intersected the cell. The conversion between $J_{\htwo}$ and $k_{\rm thick,\htwo}$ is implicitly performed through a proper choice of the normalization of the FUV cell emissivity, $j_{*,\nu,0}$.

 \begin{figure*}
 \begin{center}
\includegraphics[width=0.9\textwidth]{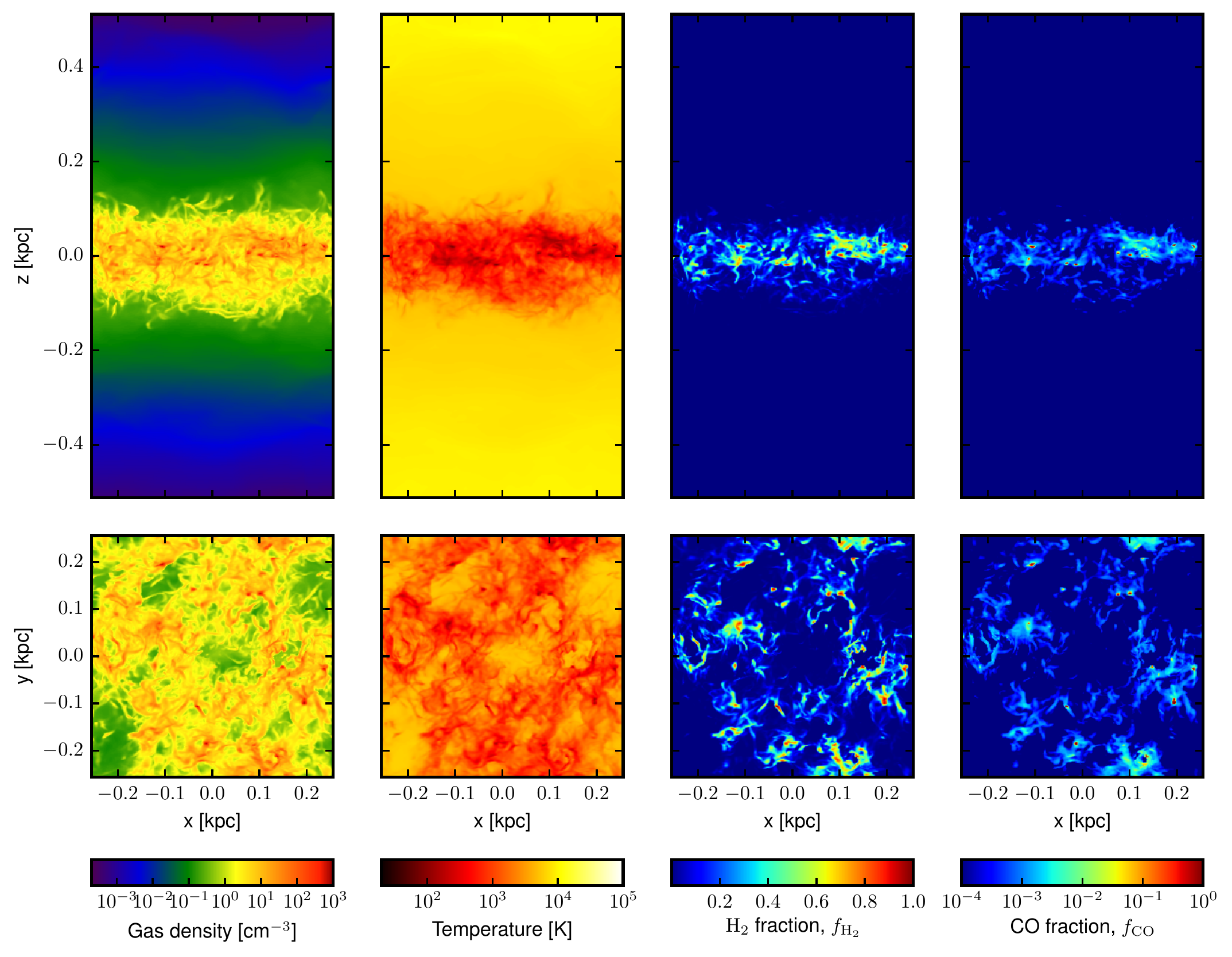}
\end{center}
\caption{Mass-weighted projections of density, temperature, $\htwo$ fraction, and CO fraction, from left-to-right, respectively. The top row are $y$-axis (disk edge on) projections while the bottom is $z$-axis (face on) projections. Note that $f_{\htwo}$ is displayed on a linear scale between $0$ and $1$, while $f_{\rm CO}$ uses a logarithmic scale.}
\label{fig:proj.F1.198.r0.iter6}
\end{figure*}

 \begin{figure*}
 \begin{center}
\includegraphics[width=1\textwidth]{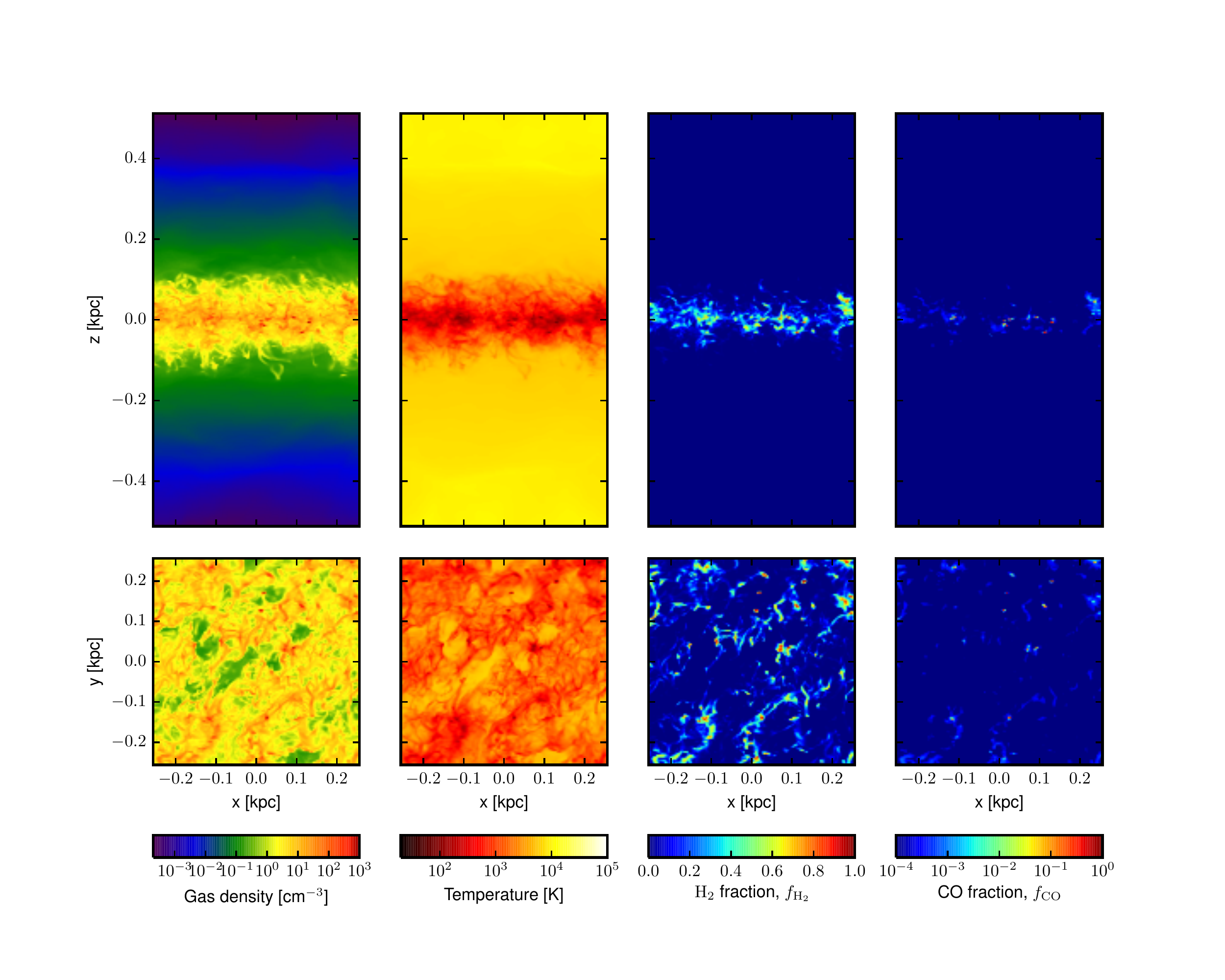}
\end{center}
\caption{Same as Figure \ref{fig:proj.F1.198.r0.iter6} but for model F2}
\label{fig:proj.F2.175.r0.iter6}
\end{figure*}

 \begin{figure*}
 \begin{center}
\includegraphics[width=1\textwidth]{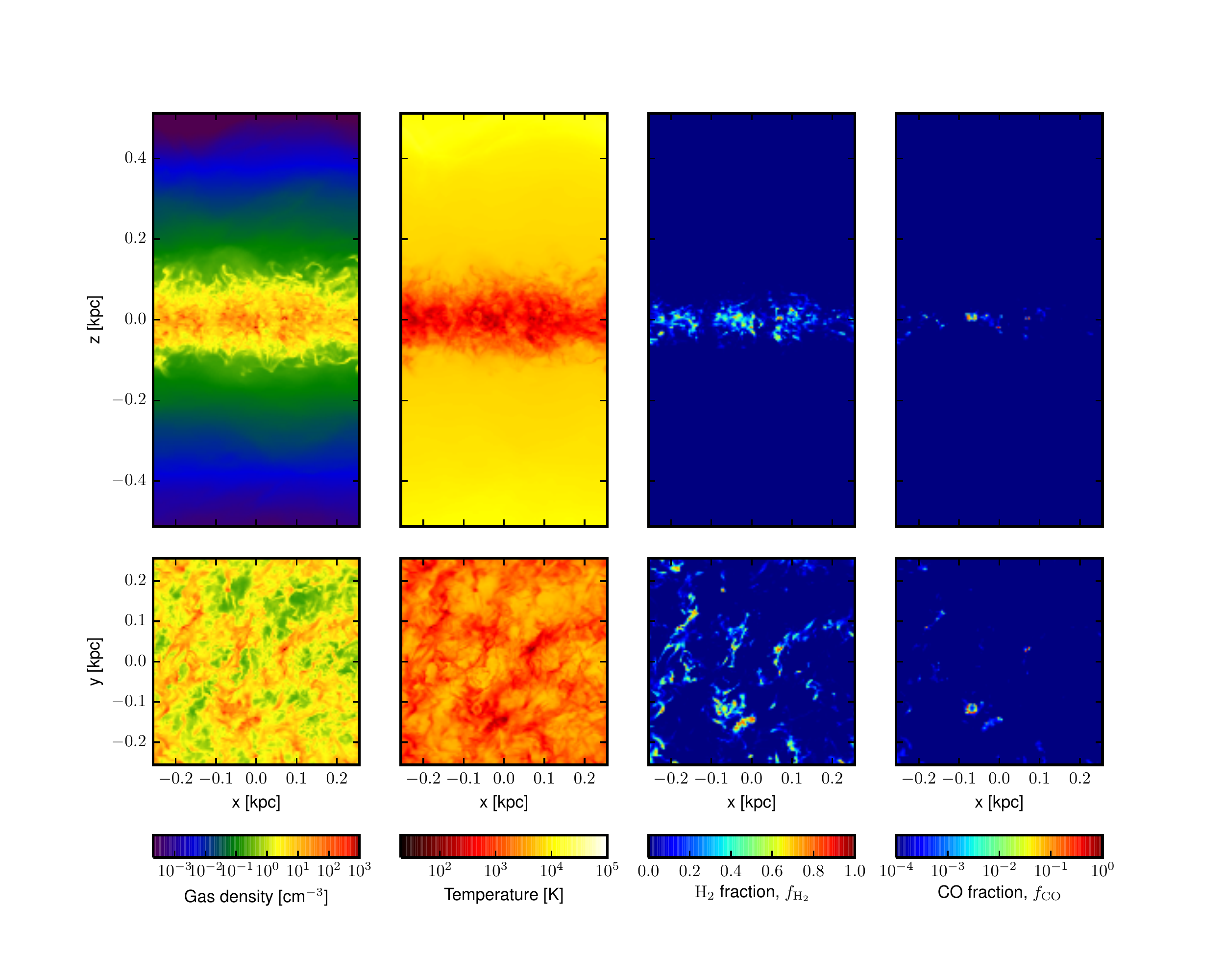}
\end{center}
\caption{Same as Figure \ref{fig:proj.F2.175.r0.iter6} but for model F3.}
\label{fig:proj.F3.234.r0.iter6}
\end{figure*}

\subsection{Local approximations for radiative shielding}
\label{sec:local}

We are interested in comparing the results obtained with the full angle-dependent ray-trace in the previous section to those produced by various local approximations for the degree of radiative shielding that have been proposed. In these approximations, the factor by which a chemical species' photodissociation rate is reduced relative to its optically thin rate, $f_{\rm shield}$, is solely dependent on an effective column density: $N_{\rm H}$ in the case of dust shielding, and $N_{\htwo} or N_{\rm \textsc{co}}$ in the case of $\htwo$ and CO shielding. Working under the approximation of uniform density and chemical composition, we can rewrite column density as the product of number density and some shielding length scale, $\{N_{\rm H},N_{\htwo}, N_{\rm \textsc{co}}\} = \{ \nh,n_{\htwo}, n_{\rm \textsc{co}} \}\times L_{\rm shield} $. This significantly reduces the complexity and computational expense of multidimensional, multifrequency radiative transfer to the relatively simple task of computing an appropriate, physically motivated shielding length $L_{\rm shield}$.  
	
Various physically motivated expressions for $L_{\rm shield}$ have been proposed and utilized, to varying degrees of success. The Sobolev approximation, alternatively known as the large velocity gradient (LVG) approach, was originally devised by \citet{Sobolev60} to study expanding stellar envelopes, but has been used extensively in simulations to model the photodissociation of $\htwo$ \citep[e.g.,][]{Yoshida06,Greif11,Stacy13,Greif13,SafranekShrader15}. It relies on the observation that in regions with a constant velocity gradient $dv_r/dr$, a photon will see Doppler shifted absorption lines with respect to its emission frame. \citet{Sobolev60} proposed that once a photon is Doppler shifted by one thermal linewidth, the photon should be free to escape. Thus, we can define the Sobolev length to be
\begin{equation}
L_{\rm sob} = \frac{\cs}{|\nabla \cdot \vect{v}|}
\label{eq:sobolev}
\end{equation}
where $\cs$ is the local sound speed and the local velocity divergence $\nabla \cdot \vect{v}$ serves as a multidimensional analog to $dv_r/dr$.

The Jeans length
\begin{equation}
L_{\rm J} = \left(\frac{\pi \cs^2}{G\rho}\right)^{1/2}
\label{eq:jeans}
\end{equation}
is the critical length scale under which the force of gravity overwhelms outward thermal pressure allowing gravitational collapse to proceed. The Jeans length is a common approximation for $L_{\rm shield}$, with the physical justification that at the center of a gravitationally collapsing core, the radiative background will be attenuated by a column of gas with a length scale approximately equal to $L_{\rm J}$.

We also investigate a shielding length based on the local density and its gradient such that
 \begin{equation}
L_{\rm dens} = \frac{\rho}{|\nabla \rho|} \,.
\label{eq:dens}
\end{equation}
The density gradient approach has been used by \citet{Gnedin09} to account for the self-shielding of $\htwo$ from photodissociating radiation and provided reasonable accuracy in the column density range $3\times10^{20}\,{\rm cm}^2 < N_{\rm H} < 3\times10^{23}\,{\rm cm}^{2}$.

An additional method to account for radiative shielding is to utilize the results of a full ray trace run to calibrate a relationship between density and column density (or equivalently, visual extinction). We we will show in Section \ref{sec:results}, this relationship can be approximated as 
 \begin{equation}
L_{\rm powerlaw} \propto \nh^{0.42}  \,.
\label{eq:powerlaw}
\end{equation}
which holds above a density of $\sim10\,\cc$ and becomes increasingly accurate with increasing density. This method naturally has the disadvantage that it is perhaps unique to the particular physical system one is examining which detracts form its generality.

A highly simplified approach to account for radiative shielding is to assume that all shielding in a given cell is due to matter in the cell in question, such that $L_{\rm shield} = dx = 2\,$ pc. This 'single-cell' approximation will surely underestimate the degree of radiative shielding, and thus molecular chemical abundances, as compared with the fiducial ray-tracing model, allowing limits to be placed on the effectiveness of radiative shielding.

Finally, while not a strictly local approach, we perform a model where ray tracing is only performed over the six directions aligned with the Cartesian axes, instead of the $48$ \textsc{healpix} selected directions in the ray trace models. This approach, known as the six-ray approximation, can be computationally efficient since the transport of rays exploits the alignment of the grid. It is a compromise between a local approach and full-fledged, multi-angle radiative transfer. 

For this to be a clean comparison, the local and ray-trace models need to utilize the same unattenuated radiation field $G_0$ so that the sole difference between the various methodologies is in the computation of $f_{\rm shield}$. To achieve this, in the local models we begin by performing the multiangle ray-trace procedure described in Section \ref{sec:raytrace}, but with all radiative shielding turned off, $f_{\rm shield}=1$, providing us with the unattenuated radiation field $G_0$ in each grid cell. This unattenuated field is then used in Equation \ref{eq:photo_shield}, along with the locally determined $f_{\rm shield}$, to compute the chemical photodissociation rates.

 \begin{figure}
 \begin{center}
\includegraphics[width=0.48\textwidth]{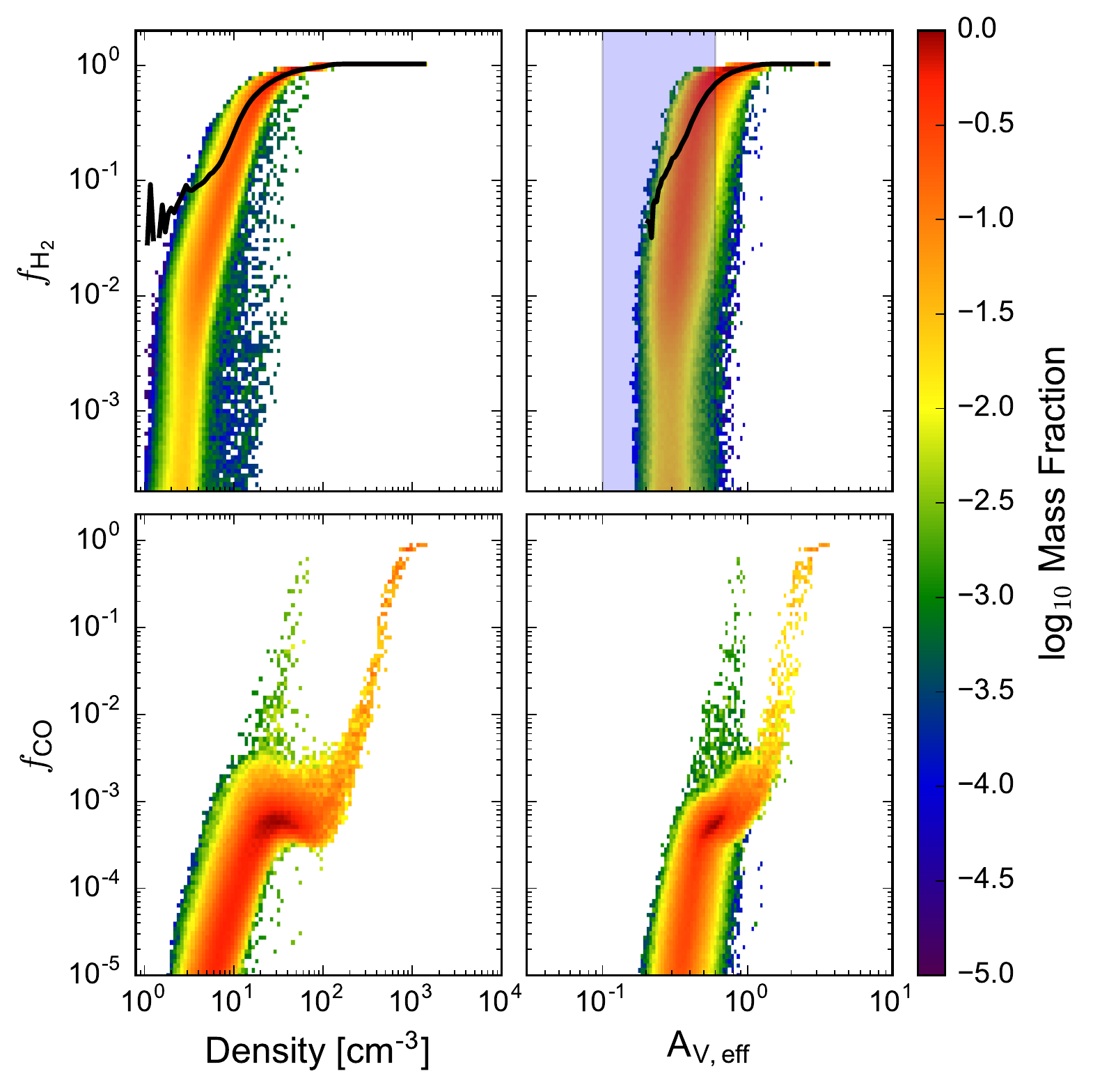}
\end{center}
\caption{Phase plots of molecular hydrogen fraction versus density (top-left), molecular hydrogen fraction versus effective visual magnitude ($A_{\rm V,eff}$, top-right), CO fraction versus density (bottom left), and CO fraction versus $A_{\rm V,eff}$ (bottom-right). Color corresponds to the mass-fraction that particular region of phase space. For comparison, the overplotted black lines are approximate analytic expressions for the $\htwo$ fraction \citep{Krumholz09,McKee10}, mass-weighted in density (left panel) or $A_{\rm V, eff}$ (right panel) bins. The shaded region in the top-right panel denotes the analytic prediction from \citet[][Equation 40]{Sternberg14a} for the maximum visual extinction before the H\,$\textsc{i}$-$\htwo$ transition, for densities from $1$ to $10\,\cc$.}
\label{fig:xco.xh2.dens.av.phase.m1.198.r0.iter6}
\end{figure}

\begin{table*}
\caption{Name and description for each model.}
\label{tab:models}
\begin{tabular}{l c  c  c  c  c c c c}
\hline\hline

Model Name  & H  &  \multicolumn{3}{|c|}{Radiative shielding$^1$}  & Midplane $G_0$ & $\zeta^2$ & Temp & RT Method \\
                       & [pc] & H$_2$ & CO & Dust &                                                                   & [s$^{-1}$H$^{-1}$] &  &  \\

\hline\hline
Fiducial \\
F1& 100 & \checkmark & \checkmark & \checkmark & 0.36 & $10^{-17}$ & Fixed & Ray Trace \\
F2 & 100 & \checkmark & \checkmark & \checkmark & 0.39 & $10^{-17}$ & Fixed & Ray Trace \\
F3  & 100 & \checkmark & \checkmark & \checkmark & 0.65 & $10^{-17}$ & Fixed & Ray Trace \\
\hline
Variation \\
V1  & 20 & \checkmark & \checkmark & \checkmark & 0.36 & $10^{-17}$ & Fixed & Ray Trace \\
V2   & 200 & \checkmark & \checkmark & \checkmark & 0.36 & $10^{-17}$ & Fixed & Ray Trace \\
V3  & 100 &  &  & \checkmark & 0.36 & $10^{-17}$ & Fixed & Ray Trace \\
V4 & 100 &  & \checkmark & \checkmark & 0.36 & $10^{-17}$ & Fixed & Ray Trace \\
V5 & 100 & \checkmark &  & \checkmark & 0.36 & $10^{-17}$ & Fixed & Ray Trace \\
V6 & 100 & \checkmark & \checkmark &  & 0.36 & $10^{-17}$ & Fixed & Ray Trace \\
V7  & 100 & \checkmark & \checkmark & \checkmark & 3.6 & $10^{-17}$ & Fixed & Ray Trace \\
V8  & 100 & \checkmark & \checkmark & \checkmark & 0.036 & $10^{-17}$ & Fixed & Ray Trace \\
V9  & 100 & \checkmark & \checkmark & \checkmark & 0.36 & $10^{-16}$ & Fixed & Ray Trace \\
V10 & 100 &  &  &  & 0.36 & $10^{-17}$ & Fixed & Ray Trace \\
\hline
Temperature \\
T1 & 100 & \checkmark & \checkmark & \checkmark & 0.36 & $10^{-17}$ & Evol & Ray Trace \\
T2 & 100 &  &  &  & 0.36 & $10^{-17}$ & Evol & Ray Trace \\
T3 & 100 & \checkmark & \checkmark & \checkmark & 3.6 & $10^{-17}$ & Evol & Ray Trace \\
T4 & 100 & \checkmark & \checkmark & \checkmark & 0.036 & $10^{-17}$ & Evol & Ray Trace \\
\hline
Local \\
L1 & 100 & \checkmark & \checkmark & \checkmark & 0.36 & $10^{-17}$ & Fixed & Jeans \\
L1a & 100 & \checkmark & \checkmark & \checkmark & 0.36 & $10^{-17}$ & Fixed & Jeans w/ T ceiling \\
L2 & 100 & \checkmark & \checkmark & \checkmark & 0.36 & $10^{-17}$ & Fixed & Sobolev \\
L3 & 100 & \checkmark & \checkmark & \checkmark & 0.36 & $10^{-17}$ & Fixed & Dens grad \\
L4 & 100 & \checkmark & \checkmark & \checkmark & 0.36 & $10^{-17}$ & Fixed & Power-law \\
L5 & 100 & \checkmark & \checkmark & \checkmark & 0.36 & $10^{-17}$ & Fixed & Single-cell \\
L6 & 100 & \checkmark & \checkmark & \checkmark & 0.36 & $10^{-17}$ & Fixed & Six-ray \\

\hline\hline
\end{tabular}

\medskip

\emph{Notes ---} (1) Radiative shielding and its subheaders denote the inclusion of a given form of radiative shielding. $\htwo$ stands for $\htwo$ self-shielding, CO denotes both CO self-shielding and cross-shielding of CO by $\htwo$, and Dust refers to the shielding of $\htwo$ and CO by dust grains. (2) $\zeta$ refers to the cosmic ray ionization rate per second per hydrogen nucleus.
\end{table*}

\section{Results}
\label{sec:results}

In our fiducial runs, we apply the method described in Section \ref{sec:raytrace} to three static snapshots from the simulations of K15. Hereafter these three fiducial runs will be labeled as F1, F2, and F3. The major difference between these snapshots is the SFR (as determined in K15 by counting the number of supernovae within the last $10$ Myr) which translates into differing normalizations of the midplane, and global, FUV intensities. Models F1, F2, and F3 have, respectively, midplane FUV intensities of $0.36$, $0.39$, and $0.65$ when normalized to the Habing field\footnote[1]{Note that these FUV intensities are $\sim3-5$ less than the interstellar average of $1.7$. This stems in part from the choice of snapshot from K15; all the snapshots selected from K15 were chosen due to the presence of large amounts of dense gas just on the verge of forming stars, and a correspondingly below average instantaneous star formation rate $\Sigma_{\rm SFR}$. Since the FUV intensity in K15 depends linearly on $\Sigma_{\rm SFR}$, this resulted in the chosen snapshots having a below average mean FUV intensity. }. All the physics described in in Sections \ref{sec:chemistry} through \ref{sec:spatial_discretize} is included, with the exception that the temperature is held constant. We additionally run models where key physical parameters are varied, or a particular physical effect is switched off, to gauge the sensitivity of our results to these variations. These models are otherwise identical to the fiducial model F1 and are labeled as follows:

 \begin{figure}
 \begin{center}
\includegraphics[width=0.5\textwidth]{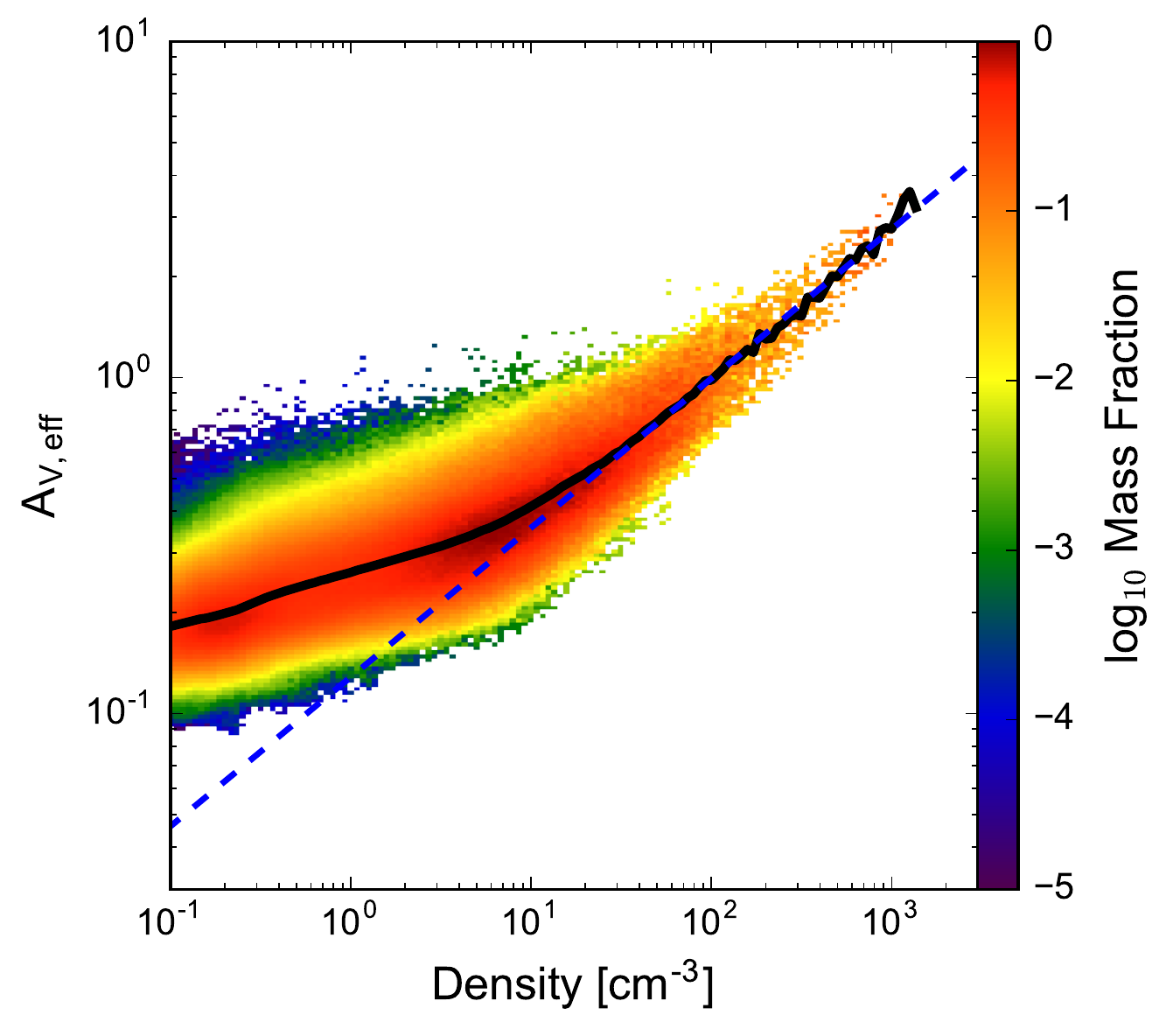}
\end{center}
\caption{Relationship between density and effective visual extinction ($A_{\rm V,eff}$, Equation \ref{eq:aveff}) in model F1. The color coding represents the mass fraction within a particular region of phase space while the black line represents the mass-weighted average. The blue dashed line is a power-law fit to the average above $\nh=10\,\cc$. Our best fitting parameters suggest $A_{\rm V,eff}\propto\nh^{0.42}$ for $\nh>10\,\cc$.}
\label{fig:dens_av_phase}
\end{figure}

\begin{enumerate}[1)]
  \item V1: Stellar density scale height decreased to $H=20\,$ pc (Equation \ref{eq:fz}), down from our fiducial value of $100\,$pc. \label{item:1}
  \item V2: Stellar density scale height increased to $200\,$ pc.
  \item V3: No self-shielding or cross-shielding for $\htwo$ or CO ($f_{\rm self,\htwo}=f_{\rm self,CO}=f_{\rm CO,\htwo} = 1$).
  \item V4: No $\htwo$ self-shielding ($f_{\rm self,\htwo} = 1$).
    \item V5: No CO self- or cross-shielding ($f_{\rm self,CO}=f_{\rm CO,\htwo} = 1$).
    \item V6: No dust shielding for $\htwo$ and CO. Dust shielding is still included for every other chemical species.
    \item V7: Normalization of the midplane FUV intensity is increased by a factor of $10$.
    \item V8: Normalization of the midplane FUV intensity is decreased by a factor of $10$.
    \item V9: Cosmic ray ionization rate increased by a factor of 10 to $\zeta=10^{-16}\,{\rm s}^{-1}$, more consistent with recent measurements \citep[e.g.,][]{Indriolo12}.
     \item V10: All radiative shielding turned off.
\end{enumerate}

We present two models where the temperature, in addition to the chemical abundances, is evolved:

\begin{enumerate}[1)]
    \item T1: Temperature is evolved in addition to chemical abundances, otherwise identical to model F1.
    \item T2: Temperature is evolved in addition to chemical abundances, and all radiative shielding turned off.
    \item T3: Temperature is evolved in addition to chemical abundances, and the midplane FUV intensity is increased by a factor of 10, like in model V7.
     \item T3: Temperature is evolved in addition to chemical abundances, and the midplane FUV intensity is decreased by a factor of 10, like in model V8.
\end{enumerate}

  Finally, we run models where the local approximations described in Section \ref{sec:local}, instead of the ray-tracing machinery, are utilized to account for radiative shielding. We label these models as follows:
  
  \begin{enumerate}[1)]
  \item L1: $L_{\rm shield} = \lj$ (Equation \ref{eq:jeans})
  \item L1a: $L_{\rm shield} = \lj$, though with a temperature ceiling of $40\,\kelvin$ used in the evaluation of $\lj$.
  \item L2: $L_{\rm shield} = L_{\rm sobolev}$ (Equation \ref{eq:sobolev})
  \item L3: $L_{\rm shield} = L_{\rm dens}$ (Equation \ref{eq:dens})
  \item L4: $L_{\rm shield}$ calibrated to be a function of density from results of fiducial models (Equation \ref{eq:powerlaw}).
    \item L5: $L_{\rm shield} = dx=2\,$pc, the local cell size; radiative shielding is only included from a single cell.	
    \item L6: Six-ray approximation. Not strictly a local quantity, though still a commonly employed approximation to the full radiative transfer problem.
\end{enumerate}

In Table \ref{tab:models} we summarize the parameters of each model, and in Table \ref{tab:results} we provide various quantitative measures for each model, including the total mass fraction of $\htwo$, CO, and a number of comparison measures with the fiducial run F1. This table will serve as a useful global comparison between each model we consider and will be referenced often in the remainder of the text.

\subsection{Fiducial models: F1-F3}
\label{sec:fiducial}

In Figures \ref{fig:proj.F1.198.r0.iter6} through \ref{fig:proj.F3.234.r0.iter6} we show the results of models F1, F2, and F3, the fiducial runs with ray-tracing. In each of these figures the left two columns are mass-weighted projections of gas density and temperature, representing the original data of K15 unchanged by our analysis. The right two columns of Figures \ref{fig:proj.F1.198.r0.iter6}-\ref{fig:proj.F3.234.r0.iter6} are mass-weighted projections of the $\htwo$ and CO fractions, $f_{\htwo}\equiv1-f_{\rm \textsc{H\,i}}$ and  $f_{\rm CO}$, the primary quantitative outputs of this study. These fractions are defined such that an $\htwo$ fraction of unity corresponds to all hydrogen being in the form of $\htwo$ while equivalently $f_{\rm CO}=1$ corresponds to all carbon atoms being incorporated in CO molecules. As is evident in all three snapshots, the bulk of the molecular gas is located close to the galactic midplane, within a vertical distance of roughly $80\,$pc. A prerequisite for molecules to form is the availability of dense, $\nh\gtrsim10\,\cc$, gas, controlled mainly by vertical stratification of pressure and the heating rate, which together define whether thermal equilibrium admits a cold phase. In the simulations of K15, two phases are always present at the midplane (see also \citealt*{Ostriker10}), but at high latitudes the pressure is too low for the formation of a cold phase. Morphologically, $\htwo$ is found predominantly along dense, filamentary structures that formed as a result of supernovae-driven turbulence in the parent simulation of K15. In model F1 roughly $20\,\%$ of all gas has an $\htwo$ fraction $f_{\htwo}$ exceeding $0.5$ (Table \ref{tab:results}). This $f_{\htwo}>0.5$ mass fraction drops to $15$\% and $10$\% for models F2 and F3 which have, respectively, elevated midplane FUV intensities of $G_0=0.39$ and $0.65$.

It is instructive to additionally consider a subset of the simulation volume, where the assumption of chemical equilibrium is well justified, for much of our analysis. Hereafter, we define \emph{cold, dense gas} to be that with a density $\nh>100\,\cc$ and temperature $T<100\,\kelvin$; quantitative justification for these values based on the chemical equilibrium timescale is discussed in Section \ref{sec:discussion}. In addition, these cold, dense gas clumps are more analogous to observed star forming clouds or clumps \citep[e.g.,][]{Bergin07}. By mass, approximately $3$\% ($2$\%) of gas within models F1 and F2 (F3) can be classified as cold, dense in which $100$\% of gas has $f_{\htwo}>0.5$, with corresponding average mass fractions of $f_{\htwo}=0.74$ for models F1-F3. It is important to note that the $\htwo$ fraction in low and moderate-density regions is likely an overestimate since it is computed under the assumption of chemical equilibrium, a simplification we discuss further in Section \ref{sec:discussion}. 

Compared to $\htwo$, the global CO fractions are significantly smaller in models F1, F2, and F3. In models F1 and F2, the total mass fraction of gas with a CO fraction $f_{\rm CO}>0.5$ is $\approx.3$\% and three times less for model F3. In cold, dense gas this fraction increases to $\approx10$\% for models F1 and F2 and $6$\% for F3. Significant fractions of CO ($f_{\rm CO}>0.1$) are only marginally visible in Figures \ref{fig:proj.F1.198.r0.iter6}-\ref{fig:proj.F3.234.r0.iter6} (note the differing scale between the $\htwo$ and CO projections) and are restricted to the densest gas ($\nh\gtrsim {\rm few}\times10^2\,\cc$) in small ($\sim5-10\,$pc), clumpy structures. Diffuse CO (defined here as $0.1>f_{\rm CO}>10^{-4}$) is slightly more widespread and tends to trace the distribution of $\htwo$ and dense gas.

We explore the relationship between the $\htwo$ and CO abundances, density, and effective visual extinction in Figure \ref{fig:xco.xh2.dens.av.phase.m1.198.r0.iter6} for model F1. Here we show phase plots of molecular hydrogen and CO abundance as a function of gas density and effective visual extinction. Since the ray-trace models do not explicitly assign a distinct $A_{\rm V}$ to each cell (unlike models L1-L5 that utilize a local approximation), we define the effective visual extinction from dust $A_{\rm V,eff}$ to be the value that would produce an identical $G_{\rm 0,eff}$ as compared with a ray-trace model, i.e., 
\begin{equation}
A_{\rm V,eff} \equiv -{\rm ln}(f_{\rm dust}) = -{\rm ln}\left(\frac{G_{\rm 0,eff}}{G_0}\right)\,,
\label{eq:aveff}
\end{equation}
where $G_0$ is the unattenuated radiation field and $G_{\rm 0,eff}$ is the dust-shielding attenuated radiation field as defined in Equation (\ref{eq:dust_shield}) with $\gamma=1$.
 The H\,$\textsc{i}$ to $\htwo$ transition (defined when $f_{\htwo}=0.5$) occurs around a density of $20\,\cc$, though more sharply over a range of effective visual extinction from $\sim0.4-0.7$. This agrees nicely with the analytic prediction from \citet[][Equation 40]{Sternberg14a} for the total atomic hydrogen column density before the H\,$\textsc{i}$ to $\htwo$ transition, which ranges from $A_{\rm V,eff}=0.1$ to $0.6$ for densities of $10$ and $1\,\cc$, respectively.
 
  Above $\nh=10^2\,\cc$, gas is composed primarily of $\htwo$. There is less scatter and a stronger relationship between $f_{\htwo}$ and density as compared with $f_{\htwo}$ and $A_{\rm v,eff}$, in the sense that a broad range of $\htwo$ concentrations can exist at the same $A_{\rm v,eff}$, particularly below $A_{\rm v,eff}\approx0.8$. As we will argue in Section \ref{sec:variations}, this lack of correlation stems from dust shielding (which depends on the total column density) being relatively unimportant in determining the $\htwo$ abundance; instead, self-shielding is the primary mechanism that depresses the $\htwo$ photodissociation rate and determines $f_{\htwo}$, which of course depends on $N_{\htwo}$, not $N_{\rm H}.$
 
 Interior to the H\,$\textsc{i}$ to $\htwo$ transition, the C\,$\textsc{i}$\,/\,C\,$\textsc{ii}$ to CO transition occurs sharply, at a density of $\approx 6\times10^2\,\cc$ or $A_{\rm V,eff}\approx2$. There also exists a second, lower density branch towards high $f_{\rm CO}$, though with a thermally and observationally insignificant mass fraction. The bottom panels of Figure \ref{fig:xco.xh2.dens.av.phase.m1.198.r0.iter6} furthermore emphasize how little carbon has fully converted into CO. This is predominantly a result of the feedback recipe used by K15, which injects momentum (representing expanding supernova blast waves) the instant density exceeds the star formation threshold, rather than allowing any time delay, a potential limitation we discuss further in Section \ref{sec:discussion}. 

It is interesting to examine how our methodology, which integrates a chemical network to equilibrium, compares with alternative approaches where the $\htwo$ abundance is computed with relatively simple analytic expressions that depend solely on the unattenuated radiation field $G_0$, mean absorption optical depth to FUV photons $\tau\approx0.5\,A_{\rm V}$, and density $\nh$. In a series of papers, \citet{Krumholz08,Krumholz09} and \citet{McKee10} analytically studied the atomic-to-molecular transition in isotropically irradiated molecular clouds. They found the $\htwo$ fraction to be well described by the following expression:
\begin{equation}
f_{\htwo} = 1-\left(\frac{3}{4}\right)\frac{s}{1+0.25s}\,,
\label{eq:fh2_analytic}
\end{equation}
where
\begin{equation}
s=\frac{{\rm ln}(1+0.6\chi +0.01\chi^2)}{0.6\tau}\,,
\end{equation}
and
\begin{equation}
\chi=42\,G_0/\nh\,.
\label{eq:chi}
\end{equation}

In Figure \ref{fig:xco.xh2.dens.av.phase.m1.198.r0.iter6} we overplot mass-weighted averages of $f_{\htwo}$ computed with Equation \ref{eq:fh2_analytic}  on top of the $\htwo$ phase plots extracted from model F1. Above $f_{\htwo}\sim {\rm few}\times10^{-2}$, the expression from \citet{McKee10} does an impressive job at matching the results from our fiducial model, particularly when viewed as a function of density.

An additional result that can be extracted from our fiducial models is the relationship between gas density and effective visual extinction (Equation \ref{eq:aveff}), which we plot in Figure \ref{fig:dens_av_phase} for model F1. Above a density of $10\,\cc$, a rough power-law relationship begins to develop between $\nh$ and $A_{\rm V,eff}$. From $\nh=10\,\cc$ to $10^3\,\cc$, the scatter in $A_{\rm V,eff}$ at a particular density decreases by roughly a factor of two. A least-squares fit yields the relationship $A_{\rm V,eff} \approx (\nh/100\,\cc)^{0.42}$, plotted in Figure \ref{fig:dens_av_phase} as a blue-dashed line. The exact normalization ($100\,\cc$, in the previous equation) of this relationship depends on the checkpoint to which this analysis is applied, but the exponent of $0.42$ appears robust between models F1, F2, and F3. This fit is the basis for one of locally determined radiative shielding models presented in Section \ref{sec:local} (via Equation \ref{eq:powerlaw}).


\begin{table*}
\caption{Model results.}
\label{tab:results}
\begin{tabular}{llllllllll}
\hline\hline
 
 Model Name & \multicolumn{2}{c}{Mass fraction$^1$} & $M_{\htwo}/M_{\rm tot}^2$ & $M_{\rm CO}/M_{\rm C, tot}$ & $M_{\htwo}/M_{\rm \htwo, fid}$ & $M_{\rm CO}/M_{\rm CO, fid}$ & ${\rm Err}_{\htwo}^3$  &  ${\rm Err}_{\rm CO}^3$ \\
  & $f_{\htwo}>0.5$ & $f_{\rm CO}>0.5$ & & & & & & \\

\hline\hline
Fiducial \\
F1 & .194 / 1 & .325(-2) / .115 & .150 & .922(-2) & 1 & 1 &  0 & 0   \\
F2 & .152 / 1 & .348(-2) / .108 & .125 & .916(-2) & ... & ... & ... &  ...  \\
F3 & .103 / 1 & .100(-2) / .600(-1) & .862(-1) & .459(-2) & ... & ... & ... & ...  \\
\hline
Variation \\
V1 & .185 / 1 & .386(-2) / .136  & .157 & .110(-1) & 1.05 & 1.19 & .150 &  .146   \\  
V2 & .166 / 1 & .332(-2) / .118  & .138 & .920(-2) & .922 & .998 & .830(-1) &  .279(-1)  \\
V3 & .387(-2) / .137 & .254(-3) / .900(-2)  & .345(-2) & .113(-2) & .233(-1) & .119 & .739 &  .826 \\
V4 & .387(-2) / .137 & .949(-3) / .335(-1) & .337(-2) & .803(-2) & .225(-1) & .871 & .739 &  .587 \\
V5 & .151 / 1 & .456(-3) / .163(-1) & .138 & .282(-2) & .920 & .306 & .142 &  .718   \\
V6 & .896(-1) / 1 & .0 / .0 & .858(-1) & .414(-3) & .572 & .450(-1) & .429 &  .913   \\
V7 & .661(-1) / 1 & .713(-3) / .254(-1) & .540(-1) & .327(-2) & .360 & .355 & .562 &  .663   \\
V8 & .343 / 1 & .598(-2) / .182  & .253 & .544(-1) & 1.69 & 2.53 & .696 &  2.10   \\
V9 & .100 / 1 & .0 / .0 & .971(-1) & .782(-3) & .647 & .848(-1) & .353 &  3.17   \\
V10 & .0 / .0 & .0 / .0 & .166(-4) & .285(-5) & .111(-3) & .308(-3) & ... & ... \\
\hline
Temperature \\
T1 & .193 / 1 & .580(-2) / .204 & .151 & .120(-1) & 1.01 & 1.31 & .197(-1) &  .446   \\
T2 & .0 / .0 & .0 / .0 & .160(-4) & .295(-5) & .112(-3) & .320(-3) & ... &  ...  \\
T3 & .691(-1) / 1 & .713(-3) / .251(-1) & .567(-1) & .289(-2) & .377 & .313 & .549 &  .634   \\
T4 & .416 / 1 & .115(-1) / .407 & .298 & .253(-1) & 1.99 & 2.74 & .982 &  1.89   \\
\hline
Local \\
L1 & .260 / 1 & .750(-2) / .144  & .209 & .188(-1) & 1.39 & 2.04 & .390 &  1.11   \\
L1a & .235 / 1 & .410(-2) / .140  & .182 & .105(-1) & 1.21 & 1.14 & .221 &  .205   \\
L2 & .563(-1) / .983 & .651(-5) / .0 & .532(-1) & .117(-3) & .350 & .124(-1) & .617 &  .975  \\
L3 & .131 / 1  & .648(-2) / .202  & .107 & .167(-1) & .717 & 1.74 & .305 &  .876  \\
L4 & .250 / 1 & .413(-2) / .140  & .198 & .106(-1) & 1.32 & 1.15 & .316 &  .223  \\
L5 & .914(-1) / 1 & .402(-2) / .141 & .785(-1) & .820(-2) & .528 & .891 & .459 &  .257 \\
L6 & .212 / 1 & .449(-2) / .158 & .166 & .123(-1) & 1.11 & 1.34 & .110 &  .335   \\

\hline\hline
\end{tabular}

\medskip

\emph{Notes ---} Numbers in parenthesis represent the scientific notation exponent, i.e., 0.19(-3) $\equiv 0.19\times10^{-3}$. (1) The total gas mass fraction with an $\htwo$ (first column) or CO (second column) fraction greater than $0.5$. The second number, after the slash, is the equivalent quantity when only considering cold, dense gas ($\nh>100\,\cc$, $T<100\,\kelvin$). (2) The total mass in the fiducial simulation (F1) is $4.57\times10^4\,\msun$. (3) See text surrounding Equation \ref{eq:err} for the description of the $\htwo$ and CO error measurements. 
\end{table*}

 \begin{figure*}
 \begin{center}
\includegraphics[width=1\textwidth]{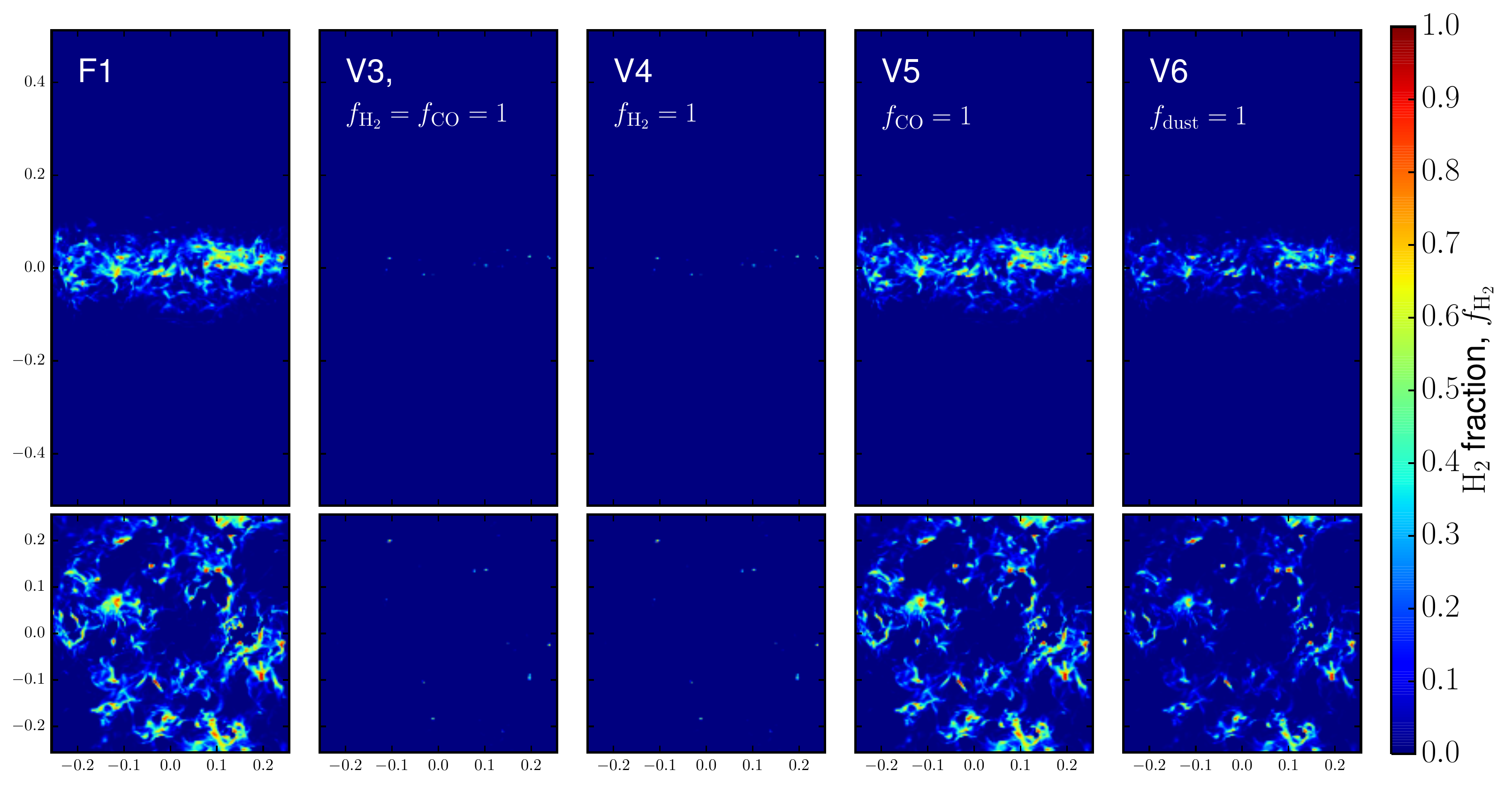}
\end{center}
\caption{Mass weighted projections of $\htwo$ exploring the role that differing types of shielding play in regulating the molecular abundances. Model V3 does not include $\htwo$ nor CO self-shielding (only dust shielding), model V4 does not include $\htwo$ self-shielding, and model V5 does not include CO self-shielding (here, we refer to the cross-shielding of CO by $\htwo$ as self-shielding). Model V6 does not include dust shielding for $\htwo$ and CO. Clearly, $\htwo$ self-shielding is necessary for a molecule rich midplane, while dense ($n\gtrsim10^3\,\cc$) clumps will be predominantly molecular regardless self-shielding operates or not. The removal of dust shielding results in a rather modest reduction of the $\htwo$ fraction. }
\label{fig:proj.shielding.h2}
\end{figure*}

 \begin{figure*}
 \begin{center}
\includegraphics[width=1\textwidth]{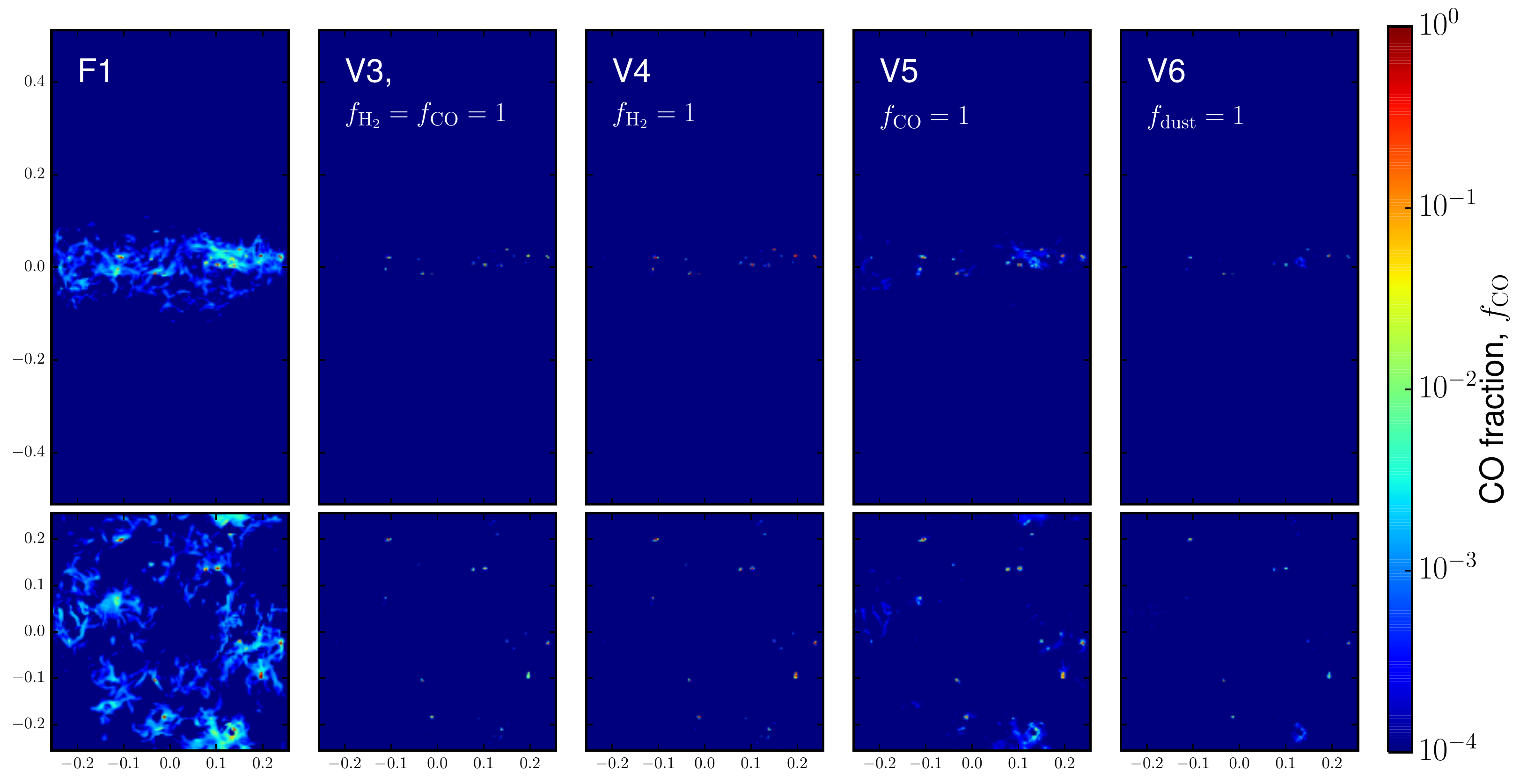}
\end{center}
\caption{Same as Figure \ref{fig:proj.shielding.h2} but for CO. The removal of any one shielding component tends to strongly depress the global CO fractions.}
\label{fig:proj.shielding.co}
\end{figure*}

\subsection{Variations on the fiducial models: V1-V10}
\label{sec:variations}

To model the spatial distribution of FUV emissivity (as required by Equation \ref{eq:soln}) we treat massive OB stars as the primary emitters of FUV radiation whose vertical density obeys the functional form of Equation \ref{eq:fz}. The fiducial model F1 utilized a vertical scale height of $H=100\,$pc, roughly consistent with the measured mean half-width of $\htwo$ in the inner Galaxy of $\simeq59\,$pc  \citep[][]{Bronfman00}. Models V1 and V2, which use scale heights of $H=20$ and $200\,$pc, respectively, demonstrate the relatively insensitivity of our results to $H$. Inspecting Table \ref{tab:results}, model V1 has $5\%$ more $\htwo$ and $\approx20\%$ more CO than F1, reasonable considering that a reduction (increase) in $H$ lowers (increases) the overall global number of FUV photodissociating photons. Model V2 has a corresponding decrease in amount of $\htwo$, though less than a $1\%$ reduction in the total CO mass as compared with F1. Morphologically, models V1 and V2 are virtually indistinguishable from F1.

Models V3 through V6, in which one or a number of shielding mechanisms are artificially switched off, are designed to assess the relative importance of self- and dust-shielding in determining the molecular abundances. Unsurprisingly, each of these models results in overall less $\htwo$ and CO mass than the comparison model F1. Models V3, V4, V5, V6 have, respectively, $2.3$, $2.3$, $92$, and $57$\% the mass of $\htwo$ as model F1. As for the CO mass, these percentages become $12$, $87$, $30$, and $4.5$\% (Table \ref{tab:results}).

Figures \ref{fig:proj.shielding.h2} and \ref{fig:proj.shielding.co} show mass-weighted projections of the $\htwo$ and CO fractions for models F1, V3, V4, V5, and V6. A comparison of models F1 with V3 and V4 in Figure \ref{fig:proj.shielding.h2} highlights the importance of $\htwo$ self-shielding in attenuating the ISRF and reducing the $\htwo$ photodissociation rate. The $\htwo$ fractions in models V3 and V4, which do not include $\htwo$ self-shielding (Equation \ref{eq:fh2}; $f_{\rm self,\htwo}=1$), are substantially suppressed (by nearly a factor of $50$ --- see Table \ref{tab:results}) compared with models that do include self-shielding. In these models, $\htwo$ fractions of unity are only realized in the highest density gas that approaches the resolution of this study. Switching off CO self- and cross-shielding, model V5, unsurprisingly has no effect on the $\htwo$ abundance.

In model V6 dust shielding is deactivated in an effort to isolate its role in reducing the direct photodissociation rates of $\htwo$ and CO. Importantly, dust shielding remains active for all other species besides $\htwo$ and CO. This choice removes the indirect effect that increased photodissociation rates for other chemical species would have in altering the chemical formation pathways of CO and, to a much lesser extent, $\htwo$. As shown in the far right panel of Figure \ref{fig:proj.shielding.h2}, the removal of dust shielding has a fairly small, though non-zero, effect on the $\htwo$ fraction. 

In contrast to $\htwo$, self-shielding, cross-shielding by $\htwo$\footnote{The CO abundance decreases with diminishing $\htwo$, though whether this stems from $\htwo$ cross-shielding or reduced CO formation rates is a subtle issue we explore later in this section}, and dust shielding are all comparably important in photoshielding CO molecules. This can be seen in Figure \ref{fig:proj.shielding.co} where each model results in a significant decrease in the amount of CO as compared with model F1. Model V4, in which $\htwo$ self-shielding is switched-off and the global $\htwo$ fractions are significantly reduced, demonstrates the importance of $\htwo$ shielding CO via their overlapping photodissociation lines in the Lyman-Werner bands. Models V5 and V6, meanwhile, respectively underscore the importance of self- and dust shielding. However, the inclusion of \emph{any} form of radiative shielding, whether from self-, cross-, or dust-shielding, results in the formation of gas with $f_{\rm CO}\approx1$ at the highest densities, $n\sim10^3\,\cc$. Model V10 (not shown), in which all radiative shielding is turned off, results in no gas with $f_{\rm CO} > 10^{-4}$, a statement that additionally applies to $\htwo$.

\begin{figure}
 \begin{center}
\includegraphics[width=0.53\textwidth]{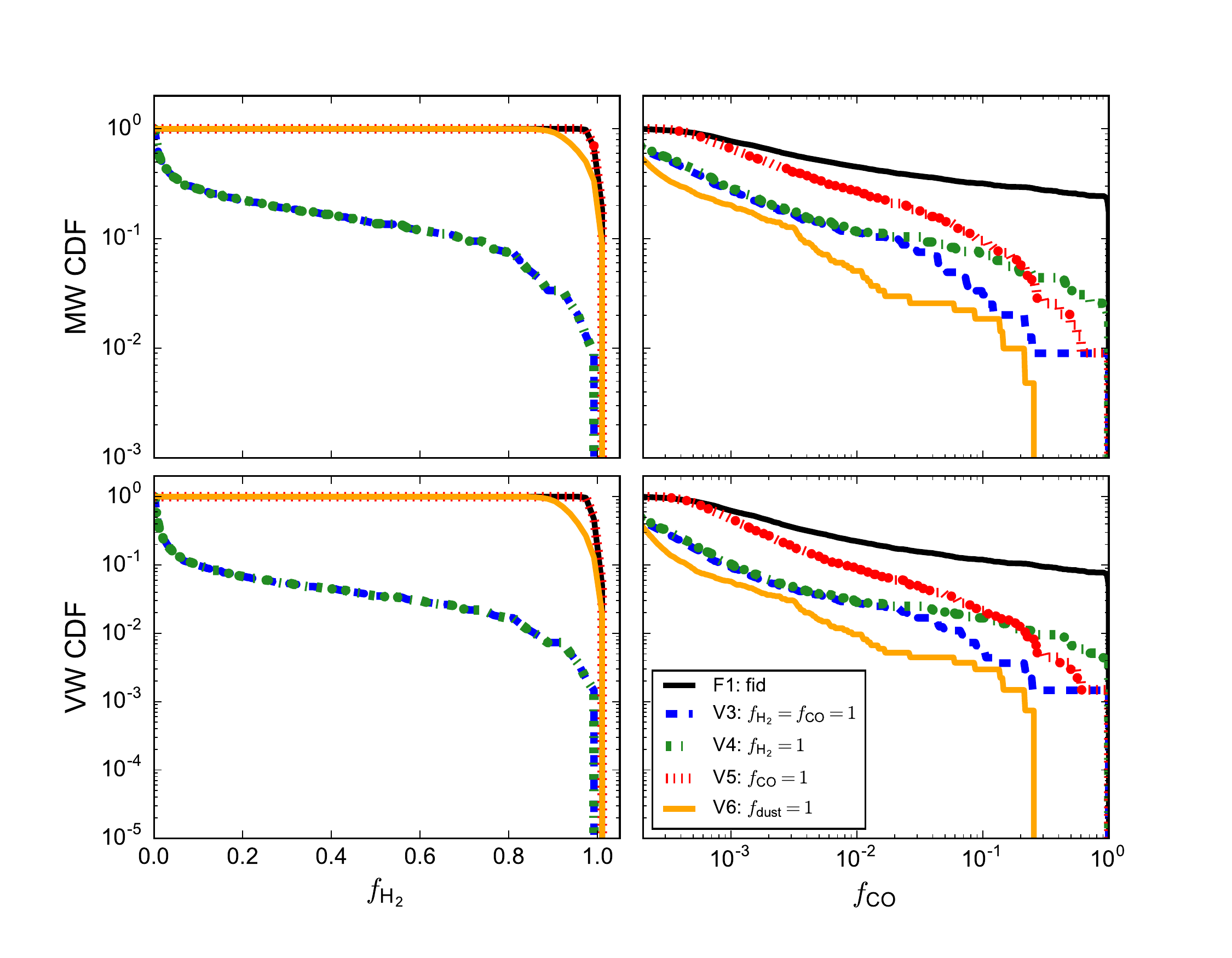}
\end{center}
\caption{Mass weighted (top row) and volume weighted (bottom row) cumulative distribution functions (CDFs) of the $\htwo$ (left panels) and CO fraction (right panels). This analysis is restricted to cold, dense gas ($\nh>100\,\cc$, $T<100\,\kelvin$). Lines of different color denote models with different treatments of dust- and self-shielding as described in the text.}
\label{fig:shielding.cdf.cd}
\end{figure}

We can explore this further by examining the cumulative distribution functions (CDF) of $f_{\htwo}$ and $f_{\rm CO}$ for models V3-V6 and F1, which we plot in Figure \ref{fig:shielding.cdf.cd}. Here, we restrict this analysis to cold, dense gas as this makes for a more potent demonstration of the competing factors at work; qualitatively identical conclusions would be drawn from the CDFs that included all gas. Focusing first on $\htwo$ (left panels), we see the removal of dust shielding only reduces the gas fraction with $f_{\htwo} \gtrsim0.9$ as compared with model F1. Turning off $\htwo$ self-shielding (models V3 and V4), conversely, suppresses the $\htwo$ mass $\sim10-50\%$ for gas with $f_{\htwo}\gtrsim0.1$. Here it is clear that by and large the dominant mechanism by which $\htwo$ is shielded from the FUV ISRF is self-shielding, at least for the range of ISRF properties probed in these simulations. In the analytic models of \citet{Krumholz08}, the relative importance of self-shielding and dust shielding depends on the value of $\chi$, Equation \ref{eq:chi}. At Milky Way values of $\chi$ the two processes are about equally important, but as noted above, the simulations we analyze here have radiation field intensities that are $\sim 2-5$ times smaller than those found in the Solar neighborhood, so the relative dominance of self-shielding is not surprising.

As for CO, a great deal of information can be extracted from a comparison of the various CDFs, shown in the right panels of Figure \ref{fig:shielding.cdf.cd}. Model V4 does not include $\htwo$ self-shielding ($f_{\rm self,\htwo}=1$) and as a consequence has a highly suppressed $\htwo$ fraction. This choice encapsulates two combined effects, one chemical and one radiative, both of which act to decrease the amount of CO. First, $f_{\rm CO,\htwo}(N_{\htwo})$ (the cross-shielding factor) is reduced as the global $\htwo$ column densities are significantly reduced. Second, the two predominant CO formation pathways (one involving OH, and the other CH) both rely on the availability of $\htwo$, thus reducing the overall speed of the chemical formation pathways that lead to CO. These effects combined result in the CO mass being suppressed by a factor of $\sim5-10$ for $f_{\rm CO}\gtrsim10^{-3}$ compared with model F1 --- the suppression increases with increasing CO fraction. From this model alone, however, is it unclear which of these two effects, chemical or radiative, is dominant.

Model V5 does not include CO self-shielding nor cross-shielding from $\htwo$. As can be seen, this has little effect in gas with a low CO fraction ($f_{\rm CO}<10^{-3}$). Evidently for this 'diffuse' CO gas, CO self-shielding and $\htwo$ cross-shielding is relatively unimportant. For CO rich gas ($f_{\rm CO}\gtrsim0.5$), the mass in CO is suppressed by a factor of $20$. Chemically, there is no suppression of the $\htwo$ abundance in this model so any reduction in $f_{\rm CO}$ is entirely due to the lack of shielding from itself and $\htwo$.


The drop in the CO fraction between models V5 and V3 is a chemical, not radiative, effect and highlights the role of $\htwo$ in the CO formation pathways. Observe that the sole difference between models V5 and V3 is that the $\htwo$ fraction in model V3 is highly suppressed due to the lack of $\htwo$ self-shielding. With regards to CO, this isolates the effect that the $\htwo$ abundance has on the efficiency of the chemical pathways that lead to CO. Quantitatively, this purely chemical effect reduces the total CO mass by $\approx60$\% (see Table \ref{tab:results}). Furthermore, at the highest CO fractions, there is no difference between model V5 and V3 in terms of the CDF, while model V4 still displays a roughly order-of-magnitude drop. This implies that a reduction in the $\htwo$ fraction (via removing $\htwo$ self-shielding) decreases the CO fraction predominantly through less $\htwo$ cross-shielding and a larger CO photodissociation rate, as opposed to a reduction in the CO formation rate.

The CO fractions realized in the absence of dust shielding (model V6) are similar to that of models V3 and V4 for $f_{\rm CO}\lesssim10^{-3}$, underscoring the importance of dust shielding for CO. In other words, the consequence of removing dust shielding for CO is roughly comparable to the combined removal of CO self-shielding, $\htwo$ cross-shielding, and decreased CO formation efficiency due to a suppressed $\htwo$ abundance. Evidently, dust shielding is critical for the formation of CO fractions approaching unity. In the absence of dust shielding, no gas is able to form with $f_{\rm CO} > 0.3$. 

Models V7 and V8 consider the scenario where the overall strength of the FUV radiative intensity is, respectively, increased or decreased by a factor of $10$. Increasing the ISRF by a factor of $10$ (V7) results in a $\approx2.7\times$ decrease in both the total $\htwo$ and CO mass, while decreasing the radiative intensity tenfold (V8) results in a factor of $1.69$ more $\htwo$ and $2.53$ times more CO mass.

In model V9, the cosmic-ray ionization rate is increased from $\zeta=10^{-17}$ to $10^{-16}\,$s$^{-1}$ to gauge its impact on molecular chemistry. This has a much more severe effect on the CO abundance than the $\htwo$ abundance. As seen in Table \ref{tab:results}, increasing $\zeta$ reduced the global $\htwo$ mass by roughly a factor of $2$, while the CO mass fell to roughly $8\%$ its value in model F1. This strong dependance of the CO abundance on the cosmic ray ionization rate is an indirect effect, stemming mainly from the cosmic ray ionization of neutral Helium and the subsequent charge-exchange destructive reaction between He$^+$ and CO.

Finally, in model V10, all radiative shielding is switched off, causing a catastrophic decrease in the abundances of $\htwo$ and CO. As can be seen in Table \ref{tab:results}, this results in zero gas with an $\htwo$ or CO fraction higher than $0.5$. The total molecular mass is also significantly reduced; the total mass of $\htwo$ and CO is $0.01\%$ and $0.03\%$, respectively, as compared to model F1. This result only underscores the importance of radiative shielding in permitting the formation of significant molecular abundances. However, as will be shown in Section \ref{sec:temperature}, the removal of all radiative shielding has a relatively small effect on the temperature structure of the gas.

\subsection{Temperature evolution: T1-T4}
\label{sec:temperature}

 \begin{figure}
 \begin{center}
\includegraphics[width=0.5\textwidth]{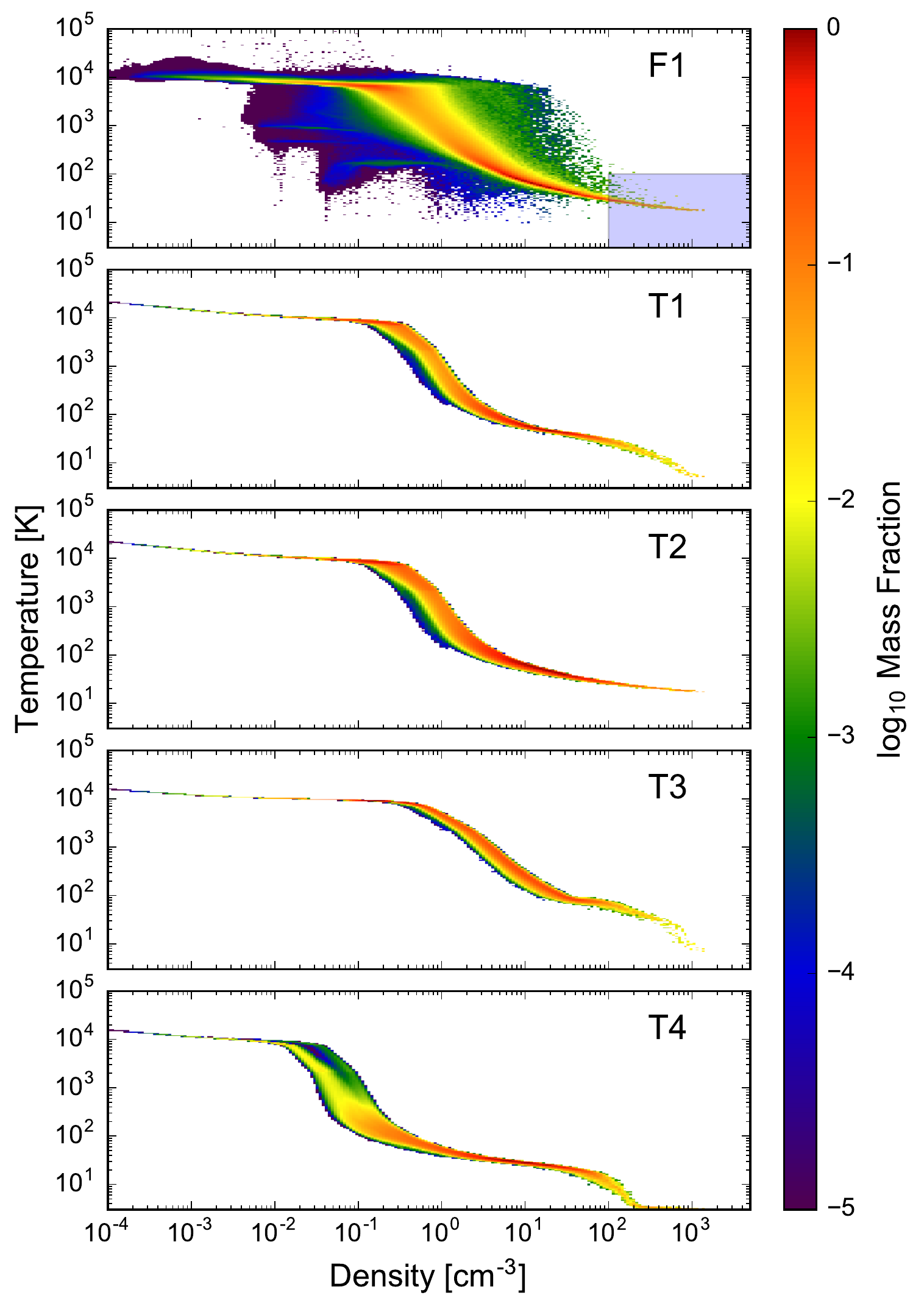}
\end{center}
\caption{Density-temperature phase diagrams for the fiducial model F1 (top), T1, T2, T3, and T4. In models T1-T4, temperature is evolved along with the chemical abundances. The density-temperature relationship of model F1, where the temperature is not evolved, is unchanged by our analysis and identical to that originally supplied by the simulation of K15. The temperature spread in model T2, mainly present at temperatures $30\,\kelvin<T<10^4\,\kelvin$ is entirely due to the vertical dependance of the unattenuated radiation field. The primary effect of changing the FUV intensity (models T3 and T4) is a shift in the density at which gas transitions from the warm to cold neutral medium. The blue shaded region in the top panel denotes this paper's definition of cold, dense gas.}
\label{fig:dts.compare}
\end{figure}

In models T1-T4, the gas temperature is evolved to equilibrium along with the chemical abundances, utilizing the thermal processes described in Section \ref{sec:thermal}. Model T1 is identical to model F1 save for the temperature evolution, while model T2 operates under optically thin conditions, i.e., no radiative shielding. In models T3 and T4 the mean FUV intensity is altered, though are otherwise identical to model T1. 

In Figure \ref{fig:dts.compare} we show density-temperature diagrams for models F1 (whose density and temperature are unchanged from the values initially supplied by K15), T1, T2, T3, and T4. At any given density, the spread in temperature for model T1 is significantly reduced as compared with model F1. In fact, any spread in temperature in model T1, at a given density, is overwhelmingly due to spatial variations in the photoelectric heating rate (Equation \ref{eq:photoelectric}) which arises due to differences in the degree of radiative shielding (via $A_{\rm v}$) and the vertical dependance of the FUV emissivity (Equation \ref{eq:fz}) and consequently the unattenuated radiation field $G_0$. Aside from this reduction in the temperature spread, the density-temperature relationship between models F1 and T1 agree remarkably well below a density of $\sim100\,\cc$. Above $\nh=10^2\,\cc$, the temperature in model T1 begins to drop below that of F1, reaching $\approx5\,\kelvin$ by $10^3\,\cc$ as a result of CO line emission, a cooling pathway not included in K15.

The relationship between density and temperature in Model T2, which does not include the effect of radiative shielding, is similar to that of model T1 below $\nh=100\,\cc$. The only significant difference between the two models is the absence of $<10\,\kelvin$ degree gas at the highest densities, due to the complete lack of any significant CO cooling, in model T2. Even by increasing the midplane $G_0$ by a factor of $10$ to $G_{\rm 0,midplane}=3.6$ (model T3), a cold phase still develops, though at a slightly larger density and equilibrium temperature of $70\,\kelvin$ before CO cooling cools the gas below $\sim10\,\kelvin$. Model T4, with its tenfold reduction in the strength of the midplane FUV intensity, exhibits the emergence of a molecular cold phase at the relatively low density of $\sim0.1\,\cc$. The molecular nature of this cold phase should be viewed as an artifact of our equilibrium assumption, given the chemical time at $\nh=0.1\,\cc$ is of order $10$ Gyr --- see Section \ref{sec:discussion}. The minimum temperature obtained in model T4 is also lower than any other model, nearly reaching the cosmic microwave background (CMB) temperature floor ($T_{\rm CMB} = 2.725\,\kelvin$) around $\nh\approx2\times10^2\,\cc$. While radiative shielding is crucial for the formation of significant molecule fractions (Section \ref{sec:variations}) and for reaching the lowest temperatures observed in the ISM, the ubiquitous appearance of a marginally cold phase (where $T\lesssim100\,\kelvin$) in all these models supports the important point that shielding is not strictly required for the transition to a cold phase, at least in the regime where the background FUV radiation field strength is relatively modest, $G_0\lesssim1$. 

\subsection{Local approximations for radiative shielding: L1-L6}
\label{sec:local}

\begin{figure*}
 \begin{center}
\includegraphics[width=1\textwidth]{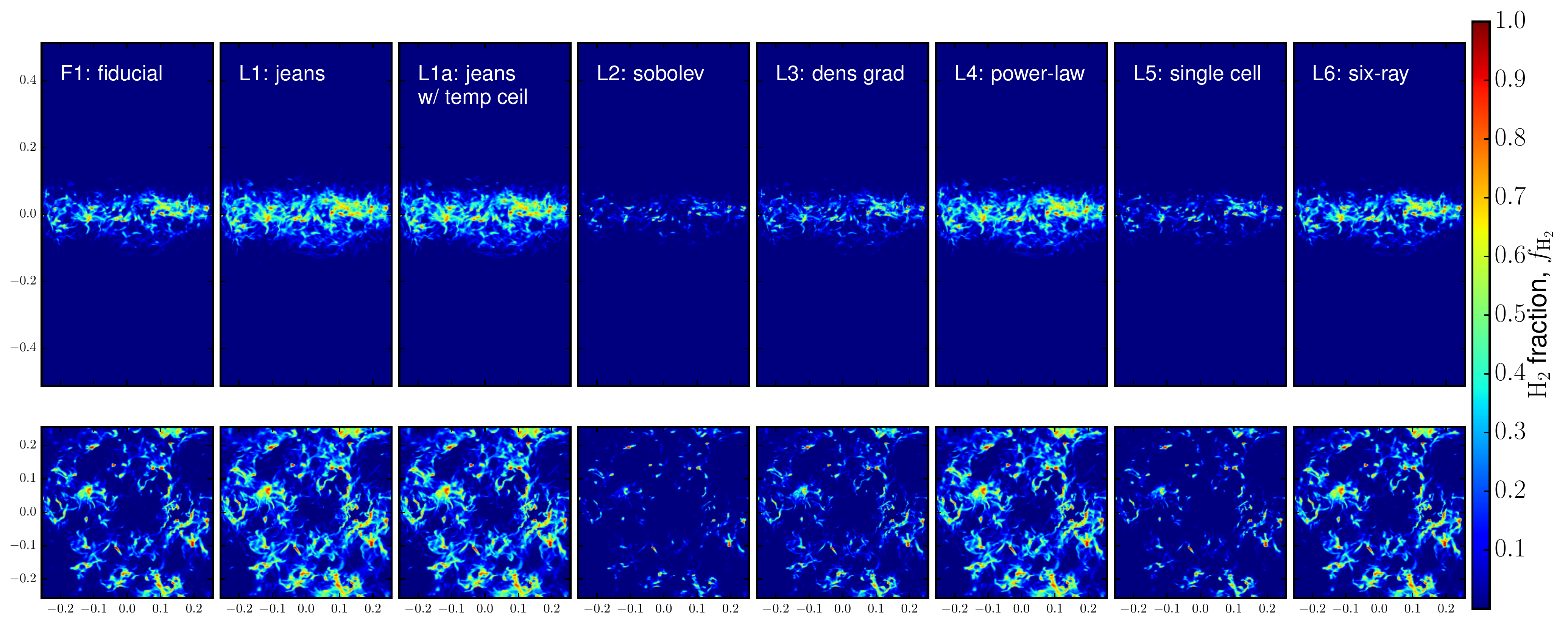}
\end{center}
\caption{Mass weighted projections of $f_{\htwo}$ for models with different locally computed approximations for the shielding length $L_{\rm shield}$.}
\label{fig:proj.local_compare.h2}
\end{figure*}

 \begin{figure*}
 \begin{center}
\includegraphics[width=1\textwidth]{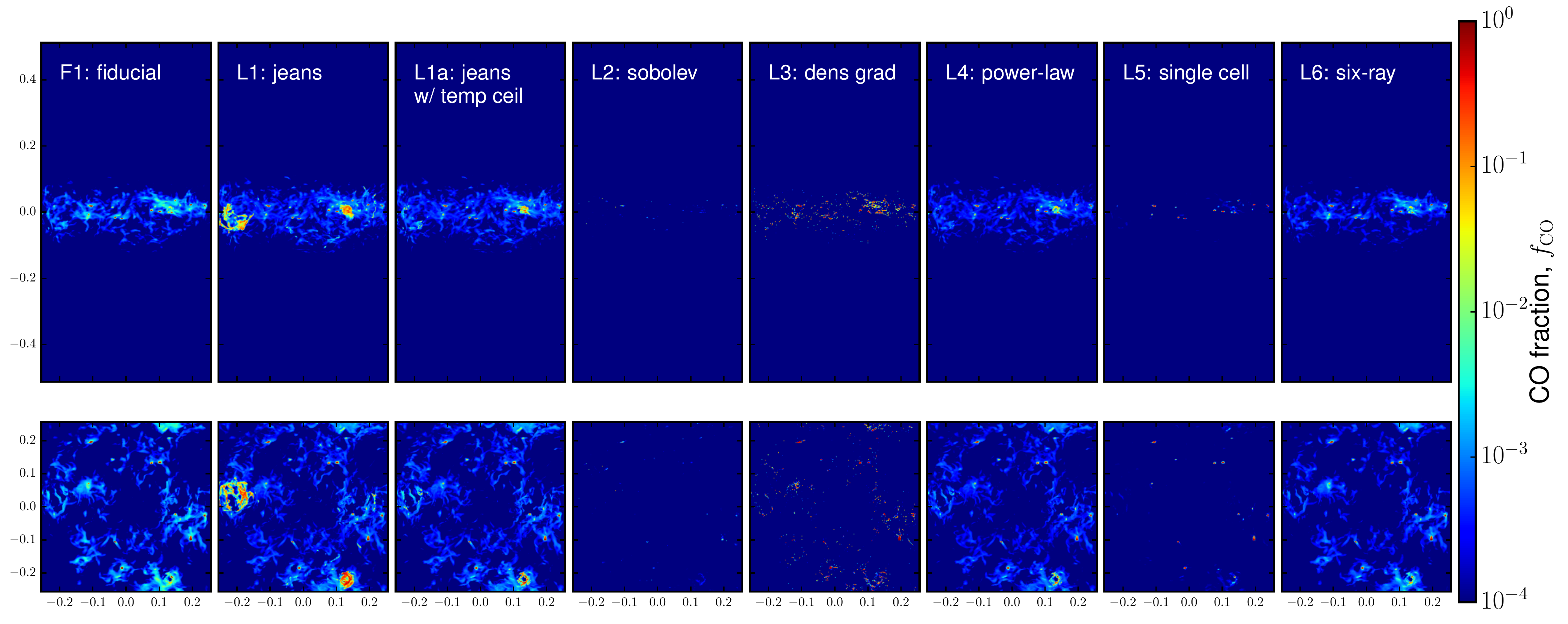}
\end{center}
\caption{Same as Figure \ref{fig:proj.local_compare.h2} but for CO.}
\label{fig:proj.local_compare.co}
\end{figure*}

The primary objective in this section is to assess the validity and accuracy of utilizing a locally computed shielding length to model radiative attenuation and thus the FUV-radiation regulated chemical state. We will do this by examining how closely the local models L1-L6 reproduce the $\htwo$ and CO abundances of the fiducial model F1 using a variety of qualitative and quantitative measures. Considering first the total mass of $\htwo$ ($M_{\htwo}$, Table \ref{tab:results} column $6$), we see all the local models come within a factor of $2$ of F1, with the exception of model L2 (Sobolev approximation) which results in $65$\% less $\htwo$. The total mass of CO is a bit more scattered than $M_{\htwo}$, though the four best performing local models (L1a, L3, L4, and L6) have $M_{\rm CO}$ within $35$\% of model F1's value. Any given local approximation model always produces an increase (or decrease) between both the total $\htwo$ and CO masses, with the exception of model L3 (density gradient) which, due to the peculiarity of using density gradients to estimate a global column density, results in $\approx30$\% less $\htwo$, and nearly $75$\% \emph{more} CO, than model F1.

The morphology and distribution of $\htwo$ between models L1-L6 and F1  (Figure \ref{fig:proj.local_compare.h2}) highlights the relative similarity between the distribution of molecular hydrogen in many of the local approximation models compared with the fiducial model. Evidently the exact prescription used to account for the attenuation of the ISRF is a relatively unimportant factor in modeling the overall distribution of $\htwo$ gas. The inclusion of \emph{some} form of radiative shielding, however, is a necessary ingredient to form any significant $\htwo$ fraction --- not shown here is model V10 (no shielding) in which no cell of gas achieves an $\htwo$ (or CO) fraction greater than $10^{-3}$ and results in a factor $\sim10^{-4}$ less $\htwo$ and CO mass.

It is important, however, to scrutinize any variation between the various local models and model F1 in Figure \ref{fig:proj.local_compare.h2}. Models in which radiative shielding is based on the local Jeans length (L1 and L1a) or a power-law fit between $\nh$ and $A_{\rm V,eff}$ (L4) appear to slightly overestimate the $\htwo$ fraction, while a shielding length computed via the velocity divergence (Sobolev length; L2), the density gradient (L3), or a single computational cell (L5), tend to underpredict $f_{\htwo}$. By eye, the best level of agreement is achieved by the six-ray approximation (model L6); this is not a strictly a local approximation, but is instead effectively the result of reducing the number of ray-tracing propagation directions from $48$ to $6$, now all aligned along the cartesian axes.

In contrast to $\htwo$, the spatial distribution and amount of CO (Figure \ref{fig:proj.local_compare.co}) shows a great deal more sensitivity to the radiative shielding prescription. As compared with model F1, models L2 and L5 fail to produce any gas with visually discernible CO. In model L3 there exists what appears to be an unphysical scattering of cells with $f_{\rm CO}$ approaching unity and a paucity of diffuse CO, $10^{-4}<f_{\rm CO}<0.1$. As with $\htwo$, the six-ray approximation (model L6) appears to do the best job in reproducing $f_{\rm CO}$. It should be noted, though, that large concentrations of CO can be difficult to discern given the small spatial extent of these concentrations.

Model L1, where $L_{\rm shield} = \lj$, seems to perform reasonably well in matching $f_{\rm CO}$ from model F1, except for the appearance of two large ($\sim50\,$pc) structures with $f_{\rm CO}>0.1$ that do not exist in model F1. We attribute this to out-of-equilibrium hot gas with density $\nh\gtrsim1\,\cc$ and temperature $T\gtrsim100\,\kelvin$. Since the Jeans length $\lj\propto T^{1/2}$, these cells have an unphysical elevated shielding length which reduces the FUV radiation intensity and boosts the CO abundance. This is an undesired effect of modeling $L_{\rm shield}$ with $\lj$ --- these cells do not represent gravitationally collapsing cores and physically should not experience a higher degree of radiative shielding. This shortcoming can be addressed quite simply, however, by introducing a temperature ceiling into the calculation of $\lj$. In model L1a the shielding length is given by this temperature capped Jeans length with an empirically chosen ceiling temperature of $T_{\rm ceiling} = 40\,\kelvin$. An inspection of Figure \ref{fig:proj.local_compare.co} confirms that these unphysical CO rich bubbles are strongly suppressed in model L1a.  

\begin{figure}
 \begin{center}
\includegraphics[width=0.42\textwidth]{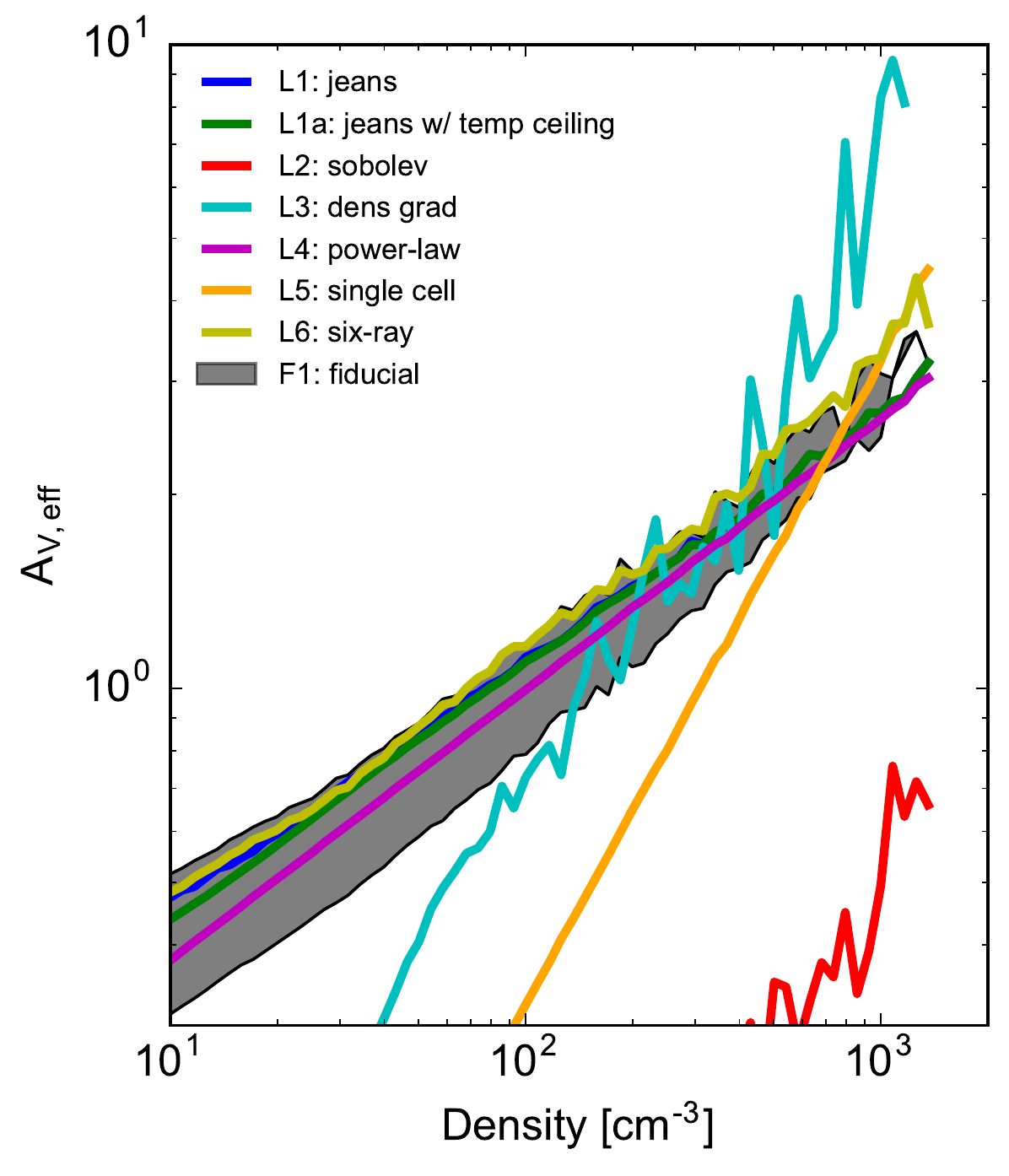}
\end{center}
\caption{Mass-weighted effective visual extinction ($A_{\rm V,eff}$, Equation \ref{eq:aveff}) vs. density. Lines of different color denote models with different treatments of radiative shielding (models L1-L6) while the black shaded region represents the $1$-$\sigma$ dispersion of $A_{\rm V,eff}$ for model F1.}
\label{fig:local_compare.dens.av.phase}
\end{figure}

To better understand the relationship amongst the local models, in Figure \ref{fig:local_compare.dens.av.phase} we plot the mass-weighted effective extinction (Equation \ref{eq:aveff}) as a function of density for each local model as compared with  model F1 (previously plotted in Figure \ref{fig:dens_av_phase}). Here it is clear how well models L1, L1a, L4, and L6 perform in matching the effective visual extinction as computed from detailed ray-tracing, at least at high densities, $\nh>10\,\cc$. Models based on local derivatives of grid-based quantities severely overpredict (L3) or underpredict (L2) the true column density. Caution should be used in interpreting Figure \ref{fig:local_compare.dens.av.phase}; in doing so, one would be led to conclude that model L4 should be the best performing local model since it was explicitly calibrated to match the $A_{\rm V,eff}-\nh$ relationship from model F1. As discussed in Section \ref{sec:variations}, $\htwo$ is primarily shielded by itself, the degree of which is a function of the $\htwo$ column density $N_{\htwo}$, not $N_{\rm H}$. Thus the $\nh-A_{\rm V,eff}$ relationship is not the ideal proxy for how well a model will predict the $\htwo$ abundance.

It is important to have a single number that characterizes the level of agreement, with regards to the spatial distribution of $\htwo$ and CO, between each local approximation model and the fiducial model F1. Many such quantitative metrics could be constructed which effectively compress a three-dimensional error distribution to a single value. For our purposes here, we favor such a metric that represents the ability of a local approximation to accurately model the atomic-to-molecular transition for both $\htwo$ and CO, rather than one that is sensitive to diffuse molecules and low density gas, since this is most relevant in regards to both observations and molecule-mediated thermo-energetics. To this end, we have found the best metric to be the mass-weighted fractional error of $\Delta \rho_x / \rho_x$ which can be expressed as
\begin{align}
{\rm Err}_x = \frac{\sum_{i,j,k}\left| \frac{\Delta \rho_x}  {\rho_{x\rm{,fid}}}\right|\rho_{x,\rm{fid}}}{\sum_{i,j,k}\rho_{x,\rm{fid}}} =  \frac{\sum_{i,j,k}|\Delta \rho_x|}{\sum_{i,j,k}\rho_{x,\rm{fid}}}
\label{eq:err}
\end{align}
where $\Delta \rho_x = \rho_{x,\rm{fid}} - \rho_{x,\rm{approx}}$, $\rho_{x,{\rm fid}}=\rho_{x,{\rm fid}}(i,j,k)$ is the density of molecule $x$ in model F1, $\rho_{x,{\rm approx}}=\rho_{x,{\rm approx}}(i,j,k)$ is the density of molecule $x$ in any separate model, $x$ denotes either $\htwo$ or CO, and the sum runs over all cells $(i,j,k)$.  Equation \ref{eq:err} is formulated such that perfect agreement between the molecular abundances would produce ${\rm Err}_x=0$, while ${\rm Err}_x=1$ signifies the local approximation model formed no molecules of species $x$ whatsoever. To prevent extremely small, negligible abundances from skewing this error metric and to better quantify the significant differences amongst the local models, we omit cells with concentrations so low where CO would not be a significant coolant, adopting a value of $10^{-2}$ for both CO and $\htwo$. We compute the above error metric between model F1 and each other model (see Table \ref{tab:models}) except for F2 and F3, whose different density maps make this error metric meaningless.

An examination of Table \ref{tab:results} shows that, amongst the local models, the six-ray approximation possesses the smallest $\htwo$ weighted error (L6, Err$_{\htwo}=0.11$), followed by the temperature capped-jeans length (L1a, $0.221$) and density gradient (L3, $0.305$). As for CO, model L1a has the lowest error (Err$_{\rm CO}=0.205$) followed by the power-law fit (L4, $0.223$) and single cell shielding (L5, $0.257$). This error metric does have limitations, and should be considered along with the other comparison measures. For instance, the relatively low Err$_{\rm CO}$ value of model L5 would suggest it does an excellent job matching the CO abundances of F1, though inspecting its CO spatial distribution (Figure \ref{fig:proj.local_compare.co}; see also Figure \ref{fig:xco_fiducial_compare}) suggests otherwise. This discrepancy is due to the CO error metric (Err$_{\rm CO}$, Equation \ref{eq:err}) being weighted by $f_{\rm CO}$, while the logarithmically scaled projections allow the detection of much more diffuse ($f_{\rm CO}\sim 10^{-3}-10^{-2}$) CO. 

\begin{figure*}
 \begin{center}
\includegraphics[width=1.0\textwidth]{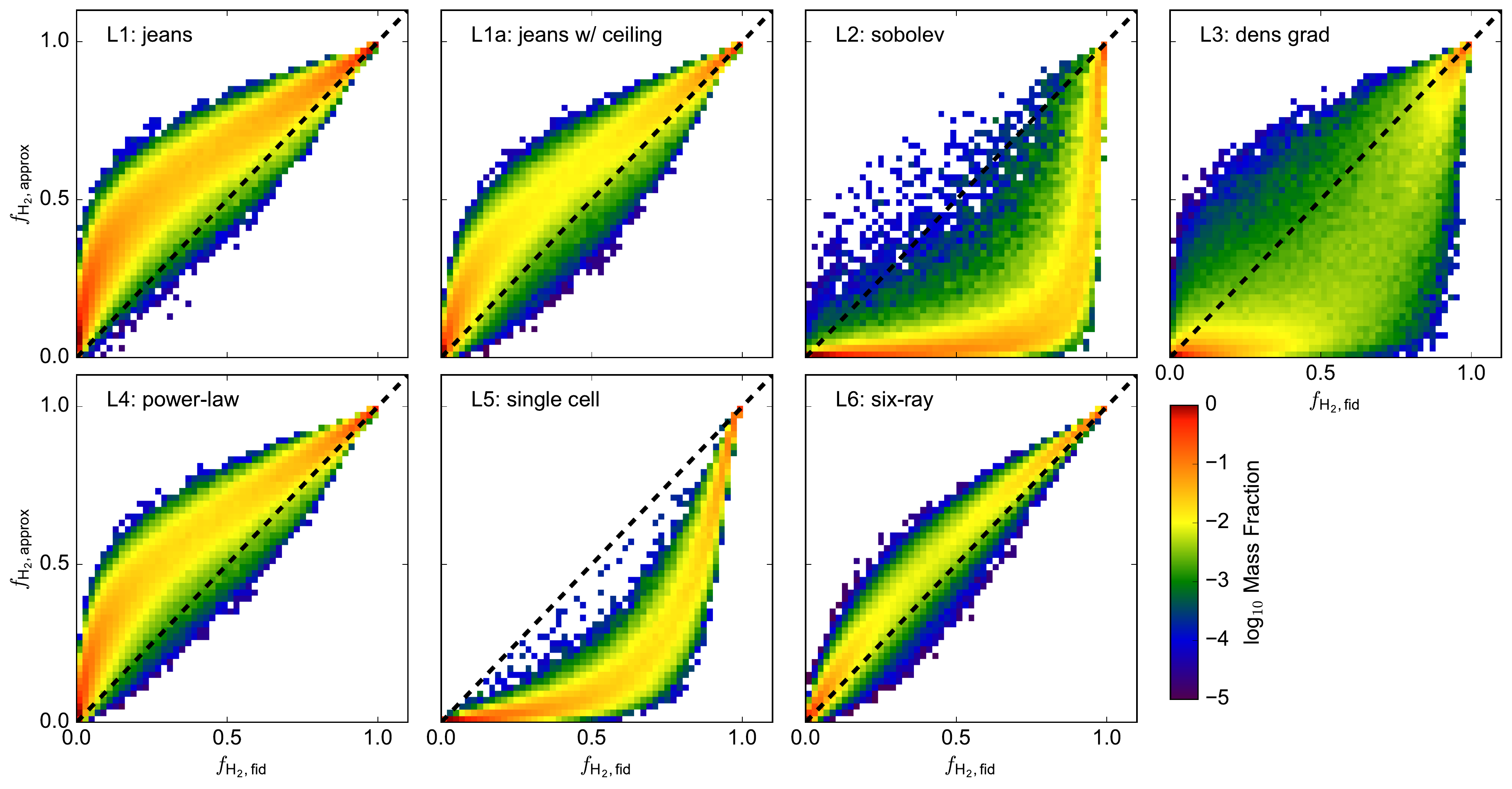}
\end{center}
\caption{Cell-by-cell, mass-weighted abundance comparisons between each local model (L1-L6) and the fiducial model F1. These plots are generated by comparing the $\htwo$ fraction between model F1 ($f_{\rm \htwo,fid}$) and each local model ($f_{\rm \htwo,approx}$) on a cell-by-cell basis. The color scale here represents the logarithm of the mass fraction in a particular region of the $f_{\rm \htwo,fid}-f_{\rm \htwo,approx}$ phase space. Perfect agreement between a local model and model F1 would exclusively lie on the diagonal dashed line defined by $f_{\rm fid}=f_{\rm approx}$. Given $f_{\htwo}$ is expressed on a linear scale, the level of agreement between the local and ray-trace models is quite good, with the largest discrepancy seen in model L2 which tends to underpredict the $\htwo$ abundance by a factor of $\sim5$.}
\label{fig:xh2_fiducial_compare}
\end{figure*}

As a final comparison between models L1-L6 and F1, we show cell-by-cell, mass-weighted abundance comparisons of $\htwo$ and CO in Figures \ref{fig:xh2_fiducial_compare} and \ref{fig:xco_fiducial_compare}. These plots are generated by comparing the $\htwo$ fraction between model F1 ($f_{\rm \htwo,fid}$) and each local model ($f_{\rm \htwo,approx}$) on a cell-by-cell basis (and similarly for CO). The color scale represents the mass fraction in a particular region of the $f_{\rm fid}-f_{\rm approx}$ phase space. Perfect agreement between a local model and model F1 would exclusively lie on the diagonal dashed line defined by $f_{\rm fid}=f_{\rm approx}$.

Local models based on the Jeans length (L1, L1a) or a power-law fit (L4) tend to slightly overpredict the $\htwo$ abundance by $\sim 20-30\%$ (see Table \ref{tab:results}). As expected, when all radiative shielding is assumed to take place in the immediate vicinity of a computational cell (model L5) the $\htwo$ abundance is slightly underpredicted by roughly a factor of $2$. Models in which the local shielding length requires a computation of a discrete gradient or divergence of a grid variable (L2 and L3) tend to exhibit a larger degree of scatter in the $\htwo$ abundance as compared to other local models (Figure \ref{fig:xh2_fiducial_compare}). As before, model L6 performs the best in matching the results of the fiducial model, overpredicting the total $\htwo$ mass by only $11\%$.

Turning to CO (Figure \ref{fig:xco_fiducial_compare}), we see a much larger scatter and discrepancy amongst the local models. Models L1 and L3 tend to significantly overpredict the CO abundance by up to four orders of magnitude. Comparing models L1 and L1a, however, shows how effectively this disagreement can be greatly reduced by imposing a temperature ceiling of $T=40\,\kelvin$ in the calculation of the Jeans, and thus shielding, length. Model L2 is unique in that effectively no gas exists with $f_{\rm CO}>0.1$. Single cell radiative shielding (model L5) results in relatively little scatter but an underprediction of the CO fraction in all cells save for those with CO fractions approaching unity. Model L5 and L1a also have a lower ${\rm Err}_{\rm CO}$ (1D error representation) than model L6, though the 2D error distribution displayed in Figure \ref{fig:xco_fiducial_compare} seems to suggest L6 to be better performing than L5 and on-par with model L1a, highlighting the utility of multiple error comparisons to gauge the effectiveness of each model in reproducing the molecular abundances of the fiducial model. 

Finally, it is interesting to note the similarity between models L1a (Jeans with temperature ceiling) and L4 (power-law fit) in their prediction of both the $\htwo$ and CO abundances. This can be attributed to their respective density scalings: Model L4 utilizes a $A_{\rm v, eff}-\nh$ relationship calibrated from model F1 ($A_{\rm v, eff}\propto \nh^{0.42}$), while the Jeans length approximation can be shown to obey a $A_{\rm v,eff}\propto \nh^{1/2}$ scaling. The similarity of these exponents should be unsurprising, though, given the physical principle underlying the Jeans length: $\lj$ is the appropriate shielding length to utilize when radiative shielding is due to surrounding material in a dense clump undergoing gravitational instability, exactly the physical entities that are able to achieve $f_{\htwo} > 0.1$ and $f_{\rm CO} > 10^{-4}$ in the simulations we analyze here.

 \begin{figure*}
 \begin{center}
\includegraphics[width=1.0\textwidth]{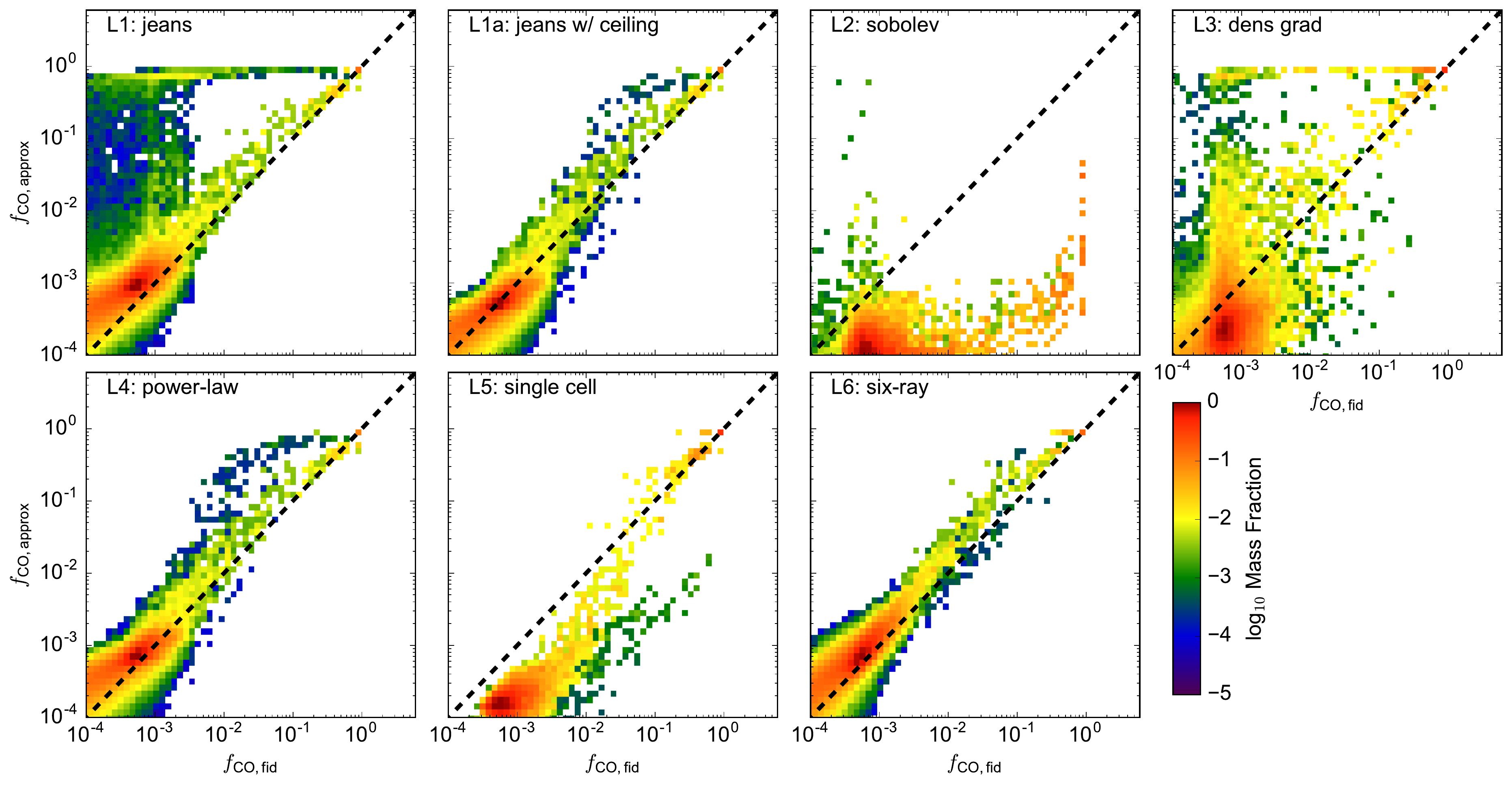}
\end{center}
\caption{Same as figure \ref{fig:xh2_fiducial_compare}, but for CO and plotted on a logarithmic scale. As compared with $\htwo$, larger discrepancies and scatters are seen when comparing the local models $f_{\rm CO}$ versus model F1. The Sobolev approximation (model L2) is unique in that effectively no gas exists with $f_{\rm CO}>0.1$. Aside from model L2, models L1 and L3 result in the largest scatters as compared with F1, in the sense that they have the capability to significantly overpredict the CO fraction. For model L1, this scatter can be effectively eliminated by imposing a temperature ceiling in the calculation of the Jeans length (model L1a).}
\label{fig:xco_fiducial_compare}
\end{figure*}

\section{Discussion and Conclusions}
\label{sec:discussion}

In this paper we have applied ray-tracing and chemical network integration to static simulation snapshots in an effort to model the distribution, morphology, and amount of $\htwo$ and CO gas in a finite, but representative, portion of a galactic disk. We have additionally performed calculations where local approximations, instead of ray-tracing, were employed to account for radiative shielding in an effort to assess the validity and accuracy of such radiative transfer alternatives.

We find significant concentrations of $\htwo$  gas ($f_{\htwo} > 0.1$) are vertically restricted to within $\approx 80\,$pc of the galactic midplane and possess a filamentary morphology which traces that of the underlying gas distribution. Gas with significant CO concentrations ($f_{\rm CO}>0.1$) is further confined to only the highest density clumps, located interior to $\htwo$ rich gas. Radiative shielding is crucial for the development of any gas with an $\htwo$ or CO fraction $>10^{-3}$. When the gas temperature is permitted to reach equilibrium along with the chemical abundances, we find little change from the K15 temperature values (except for gas at the highest densities, $n\gtrsim10^2\,\cc$, where CO cooling became effective), suggesting the bulk of the gas in K15 to exist in thermal equilibrium, and an overall insensitivity of temperature to radiative shielding.

Different physical mechanisms are responsible for the radiative shielding of $\htwo$ and CO. For $\htwo$, self-shielding is far dominant over dust shielding in reducing the $\htwo$ photodissociation rate in all regimes explored here. The dominant shielding mechanisms for CO, on the other hand, are regime dependent. More diffuse CO ($f_{\rm CO}\lesssim10^{-2}$) is shielded primarily by dust and $\htwo$ while CO rich gas ($f_{\rm CO}>0.1$) additionally depends on the presence of self-shielding by the molecule itself. However, this regime dependence for CO shielding may be sensitive to the large-scale gas distribution in this particular simulation.

We have compared a number of commonly used local approximations for radiative shielding to ray-tracing based solutions of the radiative transfer equation. Overall, it is promising how well many of the local models reproduce the $\htwo$ and CO abundances as compared with a more accurate ray-tracing based approach. One of the main objectives of this work is to inform multidimensional simulations as the validity of these local approximations to account for FUV radiative shielding. Based on our analysis, the six-ray approximation (L6), temperature-capped Jeans length (L1a), and power-law (L4) perform the best in matching the $\htwo$ and CO abundances of model F1. 

Among the approximations where the effective shielding length for a given cell (or, potentially, SPH particle) is exclusively computed from \emph{local} quantities (thus excluding model L6), the temperature-capped (at $T=40\,\kelvin$) Jeans length performed the best, an assessment based on its superiority in terms of matching the total $\htwo$ and CO masses of model F1, and possessing the smallest $\htwo$ and CO weighted errors (Equation \ref{eq:err}). We caution, however, that we have only explored one regime, that of a turbulent galactic disk, with relatively coarse ($2\,$pc) resolution. While this regime does include both the H~{\sc i}-$\htwo$ and, at the highest densities, C~{\sc ii}-CO atomic-to-molecular transitions, there are undoubtedly physical environments to which this analysis applies poorly.

Furthermore, the use of any such an approximation for $L_{\rm shield}$ critically depends on the physics a multidimensional simulation is attempting to probe. If one simply wishes to model the thermal impact of the H~{\sc i}-$\htwo$ and C~{\sc ii}-CO transitions, then coarse approximations, like those discussed above, are likely physically appropriate. On the other hand, we do not recommend such simple approximations when, for example, attempting to reproduce the detailed distribution and concentration of a large number of chemical species for observational comparison. In these cases, performing detailed, multi-frequency radiative transfer in post-processing, or utilizing advanced radiative transfer modules that operate on-the-fly, is a much more appropriate approach.

All the models presented here operate under the assumption of steady-state (equilibrium) chemistry, the validity of which can be assessed by comparing the chemical and dynamical timescales. Taking $\mathcal{R}\approx3.0\times10^{-17}\,{\rm cm}^{3}\,{\rm s}^{-1}$ to be the rate coefficient for $\htwo$ formation on grain surfaces, and $\mathcal{C}\equiv \langle\nh^2\rangle/\langle\nh\rangle^2$ to be the gas clumping factor on unresolved scales, the chemical time, the timescale for fully atomic gas to convert to molecular form, is given by $t_{\rm chem}\approx (\nh \mathcal{R}\mathcal{C} )^{-1}\approx10^3\,\nh^{-1}\mathcal{C}^{-1}\,$Myr. The dynamical time, roughly the timescale for turbulent motions to displace molecular gas, can be expressed as $t_{\rm dyn} \approx 1\,{\rm Myr}\,(L/{\rm pc})^{1/2}$, which assumes a linewidth-size relation $v_{\rm turb}\approx1\,{\rm km}\, {\rm s}^{-1}(L/{\rm pc})^{1/2}$ \citep{Dobbs14}. The assumption of equilibrium is violated when $t_{\rm dyn} < t_{\rm chem}$, or equivalently when $\nh\lesssim10^2\,\cc\,(L/{\rm pc})^{-1/2}\,(\mathcal{C}/10)^{-1}$. Thus the equilibrium chemistry is reasonably justified given we focus our analysis on $\htwo$ within the high density H~{\sc i}-H$_2$ transition. Non-equilibrium effects, however, are likely important at lower densities where the $\htwo$ formation timescale is much longer. This back-of-the-envelope argument, though, should be compared with actual findings from multidimensional simulations that included non-equilibrium chemistry. The driven turbulence simulations of \citet{Glover07b} identified moderate amounts of molecular hydrogen to be out of chemical equilibrium, a finding they attributed to turbulent transport of $\htwo$ from high- to low-density gas that occurs on a shorter timescale than the chemistry can readjust. This same type of mass transport would likely occur within other large-scale simulations. Similarly, the colliding flow simulations of \cite{Valdivia15} show the assumption of chemical equilibrium can under- or over-predict the true $\htwo$ abundance depending on density range and integration time, with the largest discrepancies occurring at early times and, interestingly, high densities.

While the assumption of chemical equilibrium likely overestimates the $\htwo$ fraction, particularly at low densities, the amount of CO is likely \emph{underestimated} due to the supernovae feedback prescription in K15. As discussed in Section \ref{sec:fiducial},  supernovae events in K15 were modeled by injecting momentum the instant density exceeds the star formation threshold, rather than allowing any time delay. For the original purposes of K15, which focused on the properties of diffuse atomic gas that is the main ISM mass reservoir, the exact timing of supernovae events was unimportant compared to the ability for the supernovae rate to change in time in response to the ISM state. However, for the present purposes, this lack of a time delay tends to suppress the formation of large, dense structures where large CO abundances would be expected. Nonetheless, since our primary objective is to understand the role of radiative shielding and numerical approximations to it in determining the chemical balance of the ISM, and not to study the chemical balance itself, this limitation of the simulations does not significantly interfere with our goals.

In future work, we plan to utilize our best performing local model, one in which the temperature-capped Jeans length is used as a proxy for $L_{\rm shield}$, to account for radiative shielding in time-dependent radiation-magneto-hydrodynamic AMR zoom-in simulations from galactic disk scales down to stellar cluster scales that will achieve significantly higher concentrations of dense, molecule rich gas than is present in the K15 snapshots. Post-processing these  snapshots with the tools developed here will further validate the utility of such a local approximation to accurately model FUV radiative shielding and photo-regulated chemistry. We additionally plan to extend the radiative transfer post-processing analysis performed here to different physical systems, notably isolated, dwarf galaxies with sub-parsec resolution and densities approaching $10^4-10^5\,\cc$ \citep[e.g.,][]{Goldbaum15}. In these targets of future analyses, the presence of more and much larger CO rich structures will permit the measurement of integrated line intensities, the X-factor, and their variation with metallicity and observation line-of-sight.

\section*{Acknowledgments}

This work was supported by NASA TCAN grant NNX-14AB52G (for CTSS, MRK, CK, ECO, JMS, SL, CFM, and RIK). RIK, MRK and CFM acknowledge support
from NASA ATP grant
NNX-13AB84G. CFM and RIK acknowledge support from NSF
grant AST-1211729. RIK acknowledges support from the US Department
of Energy at the Lawrence Livermore National Laboratory
under contract DE-AC52-07NA27344.

\footnotesize{
\bibliographystyle{mn2e_fixed}
\bibliography{complete}

\hyphenation{Post-Script Sprin-ger}
\begin{thebibliography}{81}
\expandafter\ifx\csname natexlab\endcsname\relax\def\natexlab#1{#1}\fi

\bibitem[{{Bakes} \& {Tielens}(1994)}]{Bakes94}
{Bakes} E.~L.~O., {Tielens} A.~G.~G.~M., 1994, \apj, 427, 822

\bibitem[{{Bergin} \& {Tafalla}(2007)}]{Bergin07}
{Bergin} E.~A., {Tafalla} M., 2007, \araa, 45, 339

\bibitem[{{Bialy} \& {Sternberg}(2015)}]{Bialy15a}
{Bialy} S., {Sternberg} A., 2015, \mnras, 450, 4424

\bibitem[{{Bigiel} {et~al}\mbox{.}(2008){Bigiel}, {Leroy}, {Walter}, {Brinks},
  {de Blok}, {Madore}, \& {Thornley}}]{Bigiel08}
{Bigiel} F., {Leroy} A., {Walter} F., {Brinks} E., {de Blok} W.~J.~G., {Madore}
  B., {Thornley} M.~D., 2008, \aj, 136, 2846

\bibitem[{{Black} \& {van Dishoeck}(1987)}]{Black87a}
{Black} J.~H., {van Dishoeck} E.~F., 1987, \apj, 322, 412

\bibitem[{{Bronfman} {et~al}\mbox{.}(2000){Bronfman}, {Casassus}, {May}, \&
  {Nyman}}]{Bronfman00}
{Bronfman} L., {Casassus} S., {May} J., {Nyman} L.-{\AA}., 2000, \aap, 358, 521

\bibitem[{{Browning} {et~al}\mbox{.}(2003){Browning}, {Tumlinson}, \&
  {Shull}}]{Browning03a}
{Browning} M.~K., {Tumlinson} J., {Shull} J.~M., 2003, \apj, 582, 810

\bibitem[{{Cen}(1992)}]{Cen92}
{Cen} R., 1992, \apjs, 78, 341

\bibitem[{{Chen} \& {Ostriker}(2014)}]{Chen14}
{Chen} C.-Y., {Ostriker} E.~C., 2014, \apj, 785, 69

\bibitem[{{Chen} \& {Ostriker}(2015)}]{Chen15}
{Chen} C.-Y., {Ostriker} E.~C., 2015, \apj, 810, 126

\bibitem[{{Clark} {et~al}\mbox{.}(2012){Clark}, {Glover}, \&
  {Klessen}}]{Clark12}
{Clark} P.~C., {Glover} S.~C.~O., {Klessen} R.~S., 2012, \mnras, 420, 745

\bibitem[{{Creasey} {et~al}\mbox{.}(2013){Creasey}, {Theuns}, \&
  {Bower}}]{Creasey13}
{Creasey} P., {Theuns} T., {Bower} R.~G., 2013, \mnras, 429, 1922

\bibitem[{{Davis} {et~al}\mbox{.}(2012){Davis}, {Stone}, \& {Jiang}}]{Davis12}
{Davis} S.~W., {Stone} J.~M., {Jiang} Y.-F., 2012, \apjs, 199, 9

\bibitem[{{Dobbs} {et~al}\mbox{.}(2008){Dobbs}, {Glover}, {Clark}, \&
  {Klessen}}]{Dobbs08}
{Dobbs} C.~L., {Glover} S.~C.~O., {Clark} P.~C., {Klessen} R.~S., 2008, \mnras,
  389, 1097

\bibitem[{{Dobbs} {et~al}\mbox{.}(2014){Dobbs}, {Krumholz},
  {Ballesteros-Paredes}, {Bolatto}, {Fukui}, {Heyer}, {Low}, {Ostriker}, \&
  {V{\'a}zquez-Semadeni}}]{Dobbs14}
{Dobbs} C.~L. {et~al.}, 2014, Protostars and Planets VI, 3

\bibitem[{{Draine}(1978)}]{Draine78}
{Draine} B.~T., 1978, \apjs, 36, 595

\bibitem[{{Draine}(2011)}]{Draine11}
{Draine} B.~T., 2011, {Physics of the Interstellar and Intergalactic Medium}.
  Princeton University Press: Princeton, NJ

\bibitem[{{Draine} \& {Bertoldi}(1996)}]{Draine96}
{Draine} B.~T., {Bertoldi} F., 1996, \apj, 468, 269

\bibitem[{{Fall} \& {Chandar}(2012)}]{Fall12a}
{Fall} S.~M., {Chandar} R., 2012, \apj, 752, 96

\bibitem[{{Federman} {et~al}\mbox{.}(1979){Federman}, {Glassgold}, \&
  {Kwan}}]{Federman79}
{Federman} S.~R., {Glassgold} A.~E., {Kwan} J., 1979, \apj, 227, 466

\bibitem[{{Fouesneau} {et~al}\mbox{.}(2014){Fouesneau}, {Johnson}, {Weisz},
  {Dalcanton}, {Bell}, {Bianchi}, {Caldwell}, {Gouliermis}, {Guhathakurta},
  {Kalirai}, {Larsen}, {Rix}, {Seth}, {Skillman}, \& {Williams}}]{Fouesneau14a}
{Fouesneau} M. {et~al.}, 2014, \apj, 786, 117

\bibitem[{{Gent} {et~al}\mbox{.}(2013){Gent}, {Shukurov}, {Fletcher}, {Sarson},
  \& {Mantere}}]{Gent13}
{Gent} F.~A., {Shukurov} A., {Fletcher} A., {Sarson} G.~R., {Mantere} M.~J.,
  2013, \mnras, 432, 1396

\bibitem[{{Glassgold} {et~al}\mbox{.}(2012){Glassgold}, {Galli}, \&
  {Padovani}}]{Glassgold12}
{Glassgold} A.~E., {Galli} D., {Padovani} M., 2012, \apj, 756, 157

\bibitem[{{Glover} \& {Clark}(2012{\natexlab{a}})}]{Glover12a}
{Glover} S.~C.~O., {Clark} P.~C., 2012{\natexlab{a}}, \mnras, 421, 9

\bibitem[{{Glover} \& {Clark}(2012{\natexlab{b}})}]{Glover12b}
{Glover} S.~C.~O., {Clark} P.~C., 2012{\natexlab{b}}, \mnras, 426, 377

\bibitem[{{Glover} {et~al}\mbox{.}(2010){Glover}, {Federrath}, {Mac Low}, \&
  {Klessen}}]{Glover10}
{Glover} S.~C.~O., {Federrath} C., {Mac Low} M.-M., {Klessen} R.~S., 2010,
  \mnras, 404, 2

\bibitem[{{Glover} \& {Jappsen}(2007)}]{Glover07}
{Glover} S.~C.~O., {Jappsen} A.-K., 2007, \apj, 666, 1

\bibitem[{{Glover} \& {Mac Low}(2007{\natexlab{a}})}]{Glover07a}
{Glover} S.~C.~O., {Mac Low} M.-M., 2007{\natexlab{a}}, \apjs, 169, 239

\bibitem[{{Glover} \& {Mac Low}(2007{\natexlab{b}})}]{Glover07b}
{Glover} S.~C.~O., {Mac Low} M.-M., 2007{\natexlab{b}}, \apj, 659, 1317

\bibitem[{{Gnedin} {et~al}\mbox{.}(2009){Gnedin}, {Tassis}, \&
  {Kravtsov}}]{Gnedin09}
{Gnedin} N.~Y., {Tassis} K., {Kravtsov} A.~V., 2009, \apj, 697, 55

\bibitem[{{Goldbaum} {et~al}\mbox{.}(2015){Goldbaum}, {Krumholz}, \&
  {Forbes}}]{Goldbaum15}
{Goldbaum} N.~J., {Krumholz} M.~R., {Forbes} J.~C., 2015, ArXiv e-prints,
  1510.08458

\bibitem[{{Goldsmith}(2001)}]{Goldsmith01}
{Goldsmith} P.~F., 2001, \apj, 557, 736

\bibitem[{{G{\'o}rski} {et~al}\mbox{.}(2005){G{\'o}rski}, {Hivon}, {Banday},
  {Wandelt}, {Hansen}, {Reinecke}, \& {Bartelmann}}]{Gorski05}
{G{\'o}rski} K.~M., {Hivon} E., {Banday} A.~J., {Wandelt} B.~D., {Hansen}
  F.~K., {Reinecke} M., {Bartelmann} M., 2005, \apj, 622, 759

\bibitem[{{Greif} {et~al}\mbox{.}(2013){Greif}, {Springel}, \&
  {Bromm}}]{Greif13}
{Greif} T.~H., {Springel} V., {Bromm} V., 2013, ArXiv e-prints, 1305.0823

\bibitem[{{Greif} {et~al}\mbox{.}(2011){Greif}, {Springel}, {White}, {Glover},
  {Clark}, {Smith}, {Klessen}, \& {Bromm}}]{Greif11}
{Greif} T.~H., {Springel} V., {White} S.~D.~M., {Glover} S.~C.~O., {Clark}
  P.~C., {Smith} R.~J., {Klessen} R.~S., {Bromm} V., 2011, \apj, 737, 75

\bibitem[{{Habing}(1968)}]{Habing68}
{Habing} H.~J., 1968, Bull. Ast. Inst. Neth., 19, 421

\bibitem[{{Hennebelle} \& {Iffrig}(2014)}]{Hennebelle14}
{Hennebelle} P., {Iffrig} O., 2014, \aap, 570, A81

\bibitem[{{Hill} {et~al}\mbox{.}(2012){Hill}, {Joung}, {Mac Low}, {Benjamin},
  {Haffner}, {Klingenberg}, \& {Waagan}}]{Hill12}
{Hill} A.~S., {Joung} M.~R., {Mac Low} M.-M., {Benjamin} R.~A., {Haffner}
  L.~M., {Klingenberg} C., {Waagan} K., 2012, \apj, 750, 104

\bibitem[{{Hollenbach} \& {Tielens}(1999)}]{Hollenbach99}
{Hollenbach} D.~J., {Tielens} A.~G.~G.~M., 1999, Reviews of Modern Physics, 71,
  173

\bibitem[{{Hopkins} {et~al}\mbox{.}(2011){Hopkins}, {Quataert}, \&
  {Murray}}]{Hopkins11a}
{Hopkins} P.~F., {Quataert} E., {Murray} N., 2011, \mnras, 417, 950

\bibitem[{{Indriolo} \& {McCall}(2012)}]{Indriolo12}
{Indriolo} N., {McCall} B.~J., 2012, \apj, 745, 91

\bibitem[{{Joung} \& {Mac Low}(2006)}]{Joung06}
{Joung} M.~K.~R., {Mac Low} M.-M., 2006, \apj, 653, 1266

\bibitem[{{Kennicutt} {et~al}\mbox{.}(2007){Kennicutt}, {Calzetti}, {Walter},
  {Helou}, {Hollenbach}, {Armus}, {Bendo}, {Dale}, {Draine}, {Engelbracht},
  {Gordon}, {Prescott}, {Regan}, {Thornley}, {Bot}, {Brinks}, {de Blok}, {de
  Mello}, {Meyer}, {Moustakas}, {Murphy}, {Sheth}, \& {Smith}}]{Kennicutt07}
{Kennicutt}, Jr. R.~C. {et~al.}, 2007, \apj, 671, 333

\bibitem[{{Kim} {et~al}\mbox{.}(2011){Kim}, {Kim}, \& {Ostriker}}]{Kim11}
{Kim} C.-G., {Kim} W.-T., {Ostriker} E.~C., 2011, \apj, 743, 25

\bibitem[{{Kim} \& {Ostriker}(2015{\natexlab{a}})}]{Kim15}
{Kim} C.-G., {Ostriker} E.~C., 2015{\natexlab{a}}, \apj, 802, 99

\bibitem[{{Kim} \& {Ostriker}(2015{\natexlab{b}})}]{Kim15a}
{Kim} C.-G., {Ostriker} E.~C., 2015{\natexlab{b}}, \apj, 815, 67

\bibitem[{{Kim} {et~al}\mbox{.}(2013){Kim}, {Ostriker}, \& {Kim}}]{Kim13}
{Kim} C.-G., {Ostriker} E.~C., {Kim} W.-T., 2013, \apj, 776, 1

\bibitem[{{Koyama} \& {Inutsuka}(2000)}]{Koyama00}
{Koyama} H., {Inutsuka} S.-I., 2000, \apj, 532, 980

\bibitem[{{Krumholz}(2012)}]{Krumholz12e}
{Krumholz} M.~R., 2012, \apj, 759, 9

\bibitem[{{Krumholz}(2014)}]{Krumholz14}
{Krumholz} M.~R., 2014, \mnras, 437, 1662

\bibitem[{{Krumholz} {et~al}\mbox{.}(2007){Krumholz}, {Klein}, {McKee}, \&
  {Bolstad}}]{Krumholz07b}
{Krumholz} M.~R., {Klein} R.~I., {McKee} C.~F., {Bolstad} J., 2007, \apj, 667,
  626

\bibitem[{{Krumholz} {et~al}\mbox{.}(2011){Krumholz}, {Leroy}, \&
  {McKee}}]{Krumholz11b}
{Krumholz} M.~R., {Leroy} A.~K., {McKee} C.~F., 2011, \apj, 731, 25

\bibitem[{{Krumholz} {et~al}\mbox{.}(2008){Krumholz}, {McKee}, \&
  {Tumlinson}}]{Krumholz08}
{Krumholz} M.~R., {McKee} C.~F., {Tumlinson} J., 2008, \apj, 689, 865

\bibitem[{{Krumholz} {et~al}\mbox{.}(2009){Krumholz}, {McKee}, \&
  {Tumlinson}}]{Krumholz09}
{Krumholz} M.~R., {McKee} C.~F., {Tumlinson} J., 2009, \apj, 693, 216

\bibitem[{{Lee} {et~al}\mbox{.}(1996){Lee}, {Herbst}, {Pineau des Forets},
  {Roueff}, \& {Le Bourlot}}]{Lee96}
{Lee} H.-H., {Herbst} E., {Pineau des Forets} G., {Roueff} E., {Le Bourlot} J.,
  1996, \aap, 311, 690

\bibitem[{{Leroy} {et~al}\mbox{.}(2008){Leroy}, {Walter}, {Brinks}, {Bigiel},
  {de Blok}, {Madore}, \& {Thornley}}]{Leroy08}
{Leroy} A.~K., {Walter} F., {Brinks} E., {Bigiel} F., {de Blok} W.~J.~G.,
  {Madore} B., {Thornley} M.~D., 2008, \aj, 136, 2782

\bibitem[{{Leroy} {et~al}\mbox{.}(2013){Leroy}, {Walter}, {Sandstrom},
  {Schruba}, {Munoz-Mateos}, {Bigiel}, {Bolatto}, {Brinks}, {de Blok}, {Meidt},
  {Rix}, {Rosolowsky}, {Schinnerer}, {Schuster}, \& {Usero}}]{Leroy13a}
{Leroy} A.~K. {et~al.}, 2013, \aj, 146, 19

\bibitem[{{Li} {et~al}\mbox{.}(2012){Li}, {Myers}, \& {McKee}}]{Li12}
{Li} P.~S., {Myers} A., {McKee} C.~F., 2012, \apj, 760, 33

\bibitem[{{Mac Low} \& {Glover}(2012)}]{Mac-Low12a}
{Mac Low} M.-M., {Glover} S.~C.~O., 2012, \apj, 746, 135

\bibitem[{{McKee} \& {Krumholz}(2010)}]{McKee10}
{McKee} C.~F., {Krumholz} M.~R., 2010, \apj, 709, 308

\bibitem[{{Neufeld} {et~al}\mbox{.}(1995){Neufeld}, {Lepp}, \&
  {Melnick}}]{Neufeld95}
{Neufeld} D.~A., {Lepp} S., {Melnick} G.~J., 1995, \apjs, 100, 132

\bibitem[{{Ostriker} {et~al}\mbox{.}(2010){Ostriker}, {McKee}, \&
  {Leroy}}]{Ostriker10}
{Ostriker} E.~C., {McKee} C.~F., {Leroy} A.~K., 2010, \apj, 721, 975

\bibitem[{{Rijkhorst} {et~al}\mbox{.}(2006){Rijkhorst}, {Plewa}, {Dubey}, \&
  {Mellema}}]{Rijkhorst06}
{Rijkhorst} E.-J., {Plewa} T., {Dubey} A., {Mellema} G., 2006, \aap, 452, 907

\bibitem[{{Safranek-Shrader} {et~al}\mbox{.}(2015){Safranek-Shrader},
  {Montgomery}, {Milosavljevic}, \& {Bromm}}]{SafranekShrader15}
{Safranek-Shrader} C., {Montgomery} M., {Milosavljevic} M., {Bromm} V., 2015,
  ArXiv e-prints, 1501.03212

\bibitem[{{Saintonge} {et~al}\mbox{.}(2011{\natexlab{a}}){Saintonge},
  {Kauffmann}, {Kramer}, {Tacconi}, {Buchbender}, {Catinella}, {Fabello},
  {Graci{\'a}-Carpio}, {Wang}, {Cortese}, {Fu}, {Genzel}, {Giovanelli}, {Guo},
  {Haynes}, {Heckman}, {Krumholz}, {Lemonias}, {Li}, {Moran},
  {Rodriguez-Fernandez}, {Schiminovich}, {Schuster}, \&
  {Sievers}}]{Saintonge11a}
{Saintonge} A. {et~al.}, 2011{\natexlab{a}}, \mnras, 415, 32

\bibitem[{{Saintonge} {et~al}\mbox{.}(2011{\natexlab{b}}){Saintonge},
  {Kauffmann}, {Wang}, {Kramer}, {Tacconi}, {Buchbender}, {Catinella},
  {Graci{\'a}-Carpio}, {Cortese}, {Fabello}, {Fu}, {Genzel}, {Giovanelli},
  {Guo}, {Haynes}, {Heckman}, {Krumholz}, {Lemonias}, {Li}, {Moran},
  {Rodriguez-Fernandez}, {Schiminovich}, {Schuster}, \&
  {Sievers}}]{Saintonge11b}
{Saintonge} A. {et~al.}, 2011{\natexlab{b}}, \mnras, 61

\bibitem[{{Shetty} \& {Ostriker}(2012)}]{Shetty12}
{Shetty} R., {Ostriker} E.~C., 2012, \apj, 754, 2

\bibitem[{{Smith} {et~al}\mbox{.}(2014){Smith}, {Glover}, {Clark}, {Klessen},
  \& {Springel}}]{Smith14}
{Smith} R.~J., {Glover} S.~C.~O., {Clark} P.~C., {Klessen} R.~S., {Springel}
  V., 2014, \mnras, 441, 1628

\bibitem[{{Sobolev}(1960)}]{Sobolev60}
{Sobolev} V.~V., 1960, {Moving envelopes of stars}. Cambridge: Harvard
  University Press

\bibitem[{{Stacy} {et~al}\mbox{.}(2013){Stacy}, {Greif}, {Klessen}, {Bromm}, \&
  {Loeb}}]{Stacy13}
{Stacy} A., {Greif} T.~H., {Klessen} R.~S., {Bromm} V., {Loeb} A., 2013,
  \mnras, 431, 1470

\bibitem[{{Sternberg}(1988)}]{Sternberg88}
{Sternberg} A., 1988, \apj, 332, 400

\bibitem[{{Sternberg} {et~al}\mbox{.}(2014){Sternberg}, {Le Petit}, {Roueff},
  \& {Le Bourlot}}]{Sternberg14a}
{Sternberg} A., {Le Petit} F., {Roueff} E., {Le Bourlot} J., 2014, \apj, 790,
  10

\bibitem[{{Stone} {et~al}\mbox{.}(2008){Stone}, {Gardiner}, {Teuben}, {Hawley},
  \& {Simon}}]{Stone08}
{Stone} J.~M., {Gardiner} T.~A., {Teuben} P., {Hawley} J.~F., {Simon} J.~B.,
  2008, \apjs, 178, 137

\bibitem[{{Valdivia} {et~al}\mbox{.}(2015){Valdivia}, {Hennebelle},
  {G{\'e}rin}, \& {Lesaffre}}]{Valdivia15}
{Valdivia} V., {Hennebelle} P., {G{\'e}rin} M., {Lesaffre} P., 2015, ArXiv
  e-prints, 1512.05523

\bibitem[{{van Dishoeck} \& {Black}(1986)}]{vanDishoeck86}
{van Dishoeck} E.~F., {Black} J.~H., 1986, \apjs, 62, 109

\bibitem[{{van Dishoeck} \& {Black}(1988)}]{van-Dishoeck88a}
{van Dishoeck} E.~F., {Black} J.~H., 1988, \apj, 334, 771

\bibitem[{{Walch} {et~al}\mbox{.}(2014){Walch}, {Girichidis}, {Naab}, {Gatto},
  {Glover}, {W{\"u}nsch}, {Klessen}, {Clark}, {Peters}, \&
  {Baczynski}}]{Walch14}
{Walch} S.~K. {et~al.}, 2014, ArXiv e-prints, 1412.2749

\bibitem[{{Wise} \& {Abel}(2011)}]{Wise11}
{Wise} J.~H., {Abel} T., 2011, \mnras, 414, 3458

\bibitem[{{Wolfire} {et~al}\mbox{.}(2010){Wolfire}, {Hollenbach}, \&
  {McKee}}]{Wolfire10}
{Wolfire} M.~G., {Hollenbach} D., {McKee} C.~F., 2010, \apj, 716, 1191

\bibitem[{{Wong} \& {Blitz}(2002)}]{Wong02}
{Wong} T., {Blitz} L., 2002, \apj, 569, 157

\bibitem[{{Yoshida} {et~al}\mbox{.}(2006){Yoshida}, {Omukai}, {Hernquist}, \&
  {Abel}}]{Yoshida06}
{Yoshida} N., {Omukai} K., {Hernquist} L., {Abel} T., 2006, \apj, 652, 6

\end{thebibliography}

\IfFileExists{\jobname.bbl}{}
 {\typeout{}
  \typeout{******************************************}
  \typeout{** Please run "bibtex \jobname" to optain}
  \typeout{** the bibliography and then re-run LaTeX}
  \typeout{** twice to fix the references!}
  \typeout{******************************************}
  \typeout{}
 }
}

\label{lastpage}

\end{document}